\documentclass[a4paper,12pt,english]{article}

\usepackage{cancel}
\usepackage{amsfonts,bm,amssymb,euscript,array,babel,cite}
\usepackage[all,color]{xy}
\usepackage{relsize}
\usepackage{tikz}
\usepackage{url}
\usepackage{comment}
\usepackage{amsmath,amsthm}
\usepackage{mathtools}
\usepackage[]{graphicx}
\usepackage[T1]{fontenc}
\usepackage{framed}

\usepackage{empheq}
\usepackage{multirow}
\usepackage{hyperref}

\definecolor{darkgreen}{rgb}{0.0, 0.5, 0.0}

\usepackage{color}

\makeatletter
\DeclareRobustCommand\widecheck[1]{{\mathpalette\@widecheck{#1}}}
\def\@widecheck#1#2{%
	\setbox\z@\hbox{\m@th$#1#2$}%
	\setbox\tw@\hbox{\m@th$#1%
		\widehat{%
			\vrule\@width\z@\@height\ht\z@
			\vrule\@height\z@\@width\wd\z@}$}%
	\dp\tw@-\ht\z@
	\@tempdima\ht\z@ \advance\@tempdima2\ht\tw@ \divide\@tempdima\thr@@
	\setbox\tw@\hbox{%
		\raise\@tempdima\hbox{\scalebox{1}[-1]{\lower\@tempdima\box
				\tw@}}}%
	{\ooalign{\box\tw@ \cr \box\z@}}}
\makeatother

\newcommand\ringring[1]{%
	{
		\mathop{\kern0pt #1}\limits^{
			\vbox to-1.85ex{
				\kern-2ex 
				\hbox to -1pt{\hss\normalfont\kern.1em \r{}\kern-.45em \r{}\hss}%
				\vss 
			}
		}
	}
}

\newcommand\ringringring[1]{%
	{
		\mathop{\kern0pt #1}\limits^{
			\vbox to-1.85ex{
				\kern-2ex 
				\hbox to -1pt{\hss\normalfont\kern.1em \r{}\kern-.45em\r{}\kern-.45em \r{}\hss}%
				\vss 
			}
		}
	}
}

\makeatletter
\newcommand{\doublewidetilde}[1]{{%
  \mathpalette\double@widetilde{#1}%
}}
\newcommand{\double@widetilde}[2]{%
  \sbox\z@{$\m@th#1\widetilde{#2}$}%
  \ht\z@=.9\ht\z@
  \widetilde{\box\z@}%
}
\makeatother

\newsavebox{\mysaveboxM}
\newsavebox{\mysaveboxT}

\makeatletter

\newcommand{\dd}{\mathrm{d}}

\newcommand{\w}{\wedge}

\newcommand{\be}{\begin{equation}}
\newcommand{\ee}{\end{equation}}

\newcommand{\vare}{\varepsilon}
\def\nn{\nonumber}
\def \bea{\begin{eqnarray}} 
\def\eea{\end{eqnarray}}
\def\bse{\begin{subequations}}	
\def\ese{\end{subequations}}
\def\bal{\begin{align}} 
\def\eal{\end{align}}

\def\bi{\begin{itemize}} 
\def\ei{\end{itemize}}

\def\a{\alpha}  \def\g{\gamma} \def\G{\Gamma} \def\d{\delta} \def\D{\Delta}
\def\e{\epsilon} \def\vare{\varepsilon}
   \def\k{\kappa}
  \def\m{\mu}
\def\n{\nu}    \def\r{\rho}
\def\s{\sigma} \def\S{\Sigma}

\def\one{\mbox{1 \kern-.59em {\rm l}}}

\numberwithin{equation}{section}

\begin{document}

\makeatother
\parindent=0cm
\renewcommand{\title}[1]{\vspace{10mm}\noindent{\Large{\bf #1}}\vspace{8mm}} \newcommand{\authors}[1]{\noindent{\large #1}\vspace{5mm}} \newcommand{\address}[1]{{\itshape #1\vspace{2mm}}}

\begin{titlepage}

\begin{flushright}
 RBI-ThPhys-2021-42
\end{flushright}

\begin{center}

\title{ {\large {Duality, generalized global symmetries and jet space isometries}}}

  \authors{\large Athanasios {Chatzistavrakidis}$^{a,b}$, Georgios Karagiannis$^{a}$, Arash Ranjbar$^{a}$}
 
 
  \address{ $^{a}$ Division of Theoretical Physics, Rudjer Bo\v skovi\'c Institute \\ Bijeni\v cka 54, 10000 Zagreb, Croatia 

  	$^{b}$ Erwin Schr\"odinger International Institute for Mathematics and Physics \\ 
  	Boltzmanngasse 9, A-1090, Vienna, Austria 
 
 \vskip 2mm
 
 email:{\tt \{Athanasios.Chatzistavrakidis, Georgios.Karagiannis, aranjbar\}@irb.hr}
 
 }

\vskip 1cm

\vskip 1cm

\begin{abstract}

We revisit universal features of duality in linear and nonlinear relativistic scalar and Abelian 1-form theories with single or multiple fields, which exhibit ordinary or generalized global symmetries. We show that such global symmetries can be interpreted as generalized Killing isometries on a suitable, possibly graded, target space of fields or its jet space when the theory contains higher derivatives. This is realized via a generalized sigma model perspective motivated from the fact that higher spin particles can be Nambu-Goldstone bosons of spontaneously broken generalized global symmetries. We work out in detail the 2D examples of a compact scalar and the massless Heisenberg pion fireball model and the 4D examples of Maxwell, Born-Infeld and ModMax electrodynamics. In all cases we identify the 't Hooft anomaly that obstructs the simultaneous gauging of both global symmetries and confirm the anomaly matching under duality. These results readily generalize to higher gauge theories for $p$-forms. For multifield theories we discuss the transformation of couplings under duality as two sets of Buscher rules for even or odd differential forms.  

\end{abstract}


\end{center}

\vskip 2cm

\end{titlepage}

\tableofcontents

\setcounter{footnote}{0}


\section{Introduction} 

Global symmetries play a fundamental role in theoretical physics, giving rise to conserved currents and operators charged under them. Ordinary global symmetries, such as flavour, are associated to 1-form (``vector'') currents and their charged objects are particles and local operators at a point. Recent studies, however, have revisited and highlighted the existence of higher, or
generalized, global symmetries \cite{Gaiotto:2014kfa}.\footnote{An overview of these ideas and their relation to earlier developments such as gauged linear sigma models on stacks \cite{Pantev:2005zs,Hellerman:2006zs} is found in Ref. \cite{Sharpe:2015mja}.} Their conserved currents are higher degree differential forms and
their charged objects are string/brane excitations or line/defect operators. Like their ordinary
counterpart, such generalized symmetries can be gauged. Their gauging is, however, often obstructed by the presence of 't Hooft anomalies \cite{tHooft:1979rat}---see \cite{Cordova:2018cvg} for a  recent review. These are benign anomalies, which do not render a theory physically meaningless and they vanish in absence of
background fields because the associated triangle diagrams contain global currents in all vertices
(unlike Adler-Bell-Jackiw anomalies). They provide information regarding the dynamics of strongly coupled
gauge theories and crucially they are reproduced through anomaly matching in any description of
the theory under renormalization group flow. For instance, when a given theory exhibits some duality, there is anomaly matching in the dual descriptions. This is to be expected since dualities are associated to matching of global symmetries and exchange of currents and observables and therefore the dual symmetries cannot be gauged simultaneously. 

On the other hand, dualities are intimately related to (generalized) geometry. This is most easily observed in the case of scalar field theories in two dimensions. One of the most elementary facts about string theory is that a single scalar field propagating on a circle of radius $R$ exhibits a target space (T-)duality with a dual scalar propagating on a dual circle of radius $\a'/R$. This change in the geometry, and often topology, of the target space under duality is typical in string theory. A mathematical survey relating T-duality and generalized geometry/Courant algebroids may be found in the review \cite{Bouwknegt:2010zz}.  Global momentum and winding symmetries are exchanged under this T-duality, as well as the associated momentum and winding modes. These two symmetries cannot be gauged simultaneously since there is an underlying mixed 't Hooft anomaly between them, both in the original and the dual description. 

These features are pertinent in a variety of 2D scalar field theories beyond the ordinary string sigma model. For example, the massless limit of Heisenberg's pion fireball model{\footnote{See \cite{Nastase:2005rp,Kang:2005bj} for some more recent applications.}} \cite{Heisenberg:1952zz} exhibits the same T-duality with all above characteristics, as we discuss in this paper. The reason for this can be traced in the geometric origin of the global symmetries at hand. For a single compact scalar, it is the isometry of the circular target that essentially corresponds to the global symmetry of the theory. This persists in theories with multiple scalar fields, essentially generalized nonlinear sigma models in two dimensions, whose couplings $G$ and $B$ can be identified with a Riemannian metric and a 2-form in the target and they transform under T-duality via the Buscher rules \cite{Buscher:1987sk}. It is well-known that global symmetries are tied to invariance conditions for these couplings, uncovering that the target space possesses multiple isometries---see the review \cite{Giveon:1994fu} and references therein. Although this is established for the case of $G$ and $B$ being functions of the scalar fields only, a general statement for theories with couplings depending also on their derivatives, as for instance the Heisenberg model, is lacking---see however \cite{Bonifacio:2019rpv} for a discussion about Dirac-Born-Infeld models. This is one of the gaps we would like to fill in with this paper, showing that the global symmetries in such theories correspond to jet space isometries. The extension to jet space is reasonable in view of the jet manifold formulation for Lagrangian systems \cite{Anderson:1992,sard}.

Apart from the generalization of the $R\leftrightarrow 1/R$ T-duality to multiple scalar fields and higher-derivative theories, most of its features are also shared with $p$-form gauge theories in $2p+2$ (self-dual) dimensions \cite{Gaillard:1981rj}, including nonlinear ones. This is also to be expected, since electric/magnetic duality in electromagnetism and its nonlinear extensions are already largely of this type. An Abelian gauge theory for a 1-form in four dimensions exhibits two generalized 1-form global symmetries, the electric and magnetic ones, with conserved currents given by the field strength and its Hodge dual and observables being Wilson and 't Hooft line operators, which are exchanged under the duality \cite{Banks:2010zn,Cordova:2018cvg}. Clearly these global symmetries cannot be gauged simultaneously and there is a mixed 't Hooft anomaly in the theory. Moreover, spontaneous breaking of the 1-form global symmetry leads to the interpretation of a free photon as a Nambu-Goldstone boson. 

As in the case of scalar fields in two dimensions, these features appear also in higher-derivative, nonlinear theories such as Born-Infeld \cite{Born:1934gh} and the recently constructed ModMax electrodynamics \cite{Bandos:2020jsw}, both of which exhibit duality rotations and generalized global symmetries. Moreover, this discussion can be extended to theories with multiple 1-forms, which can be interpreted as generalized sigma model-like theories with couplings $G$ and $B$  both being symmetric 2-tensors in the space of fields. This sigma model philosophy is well motivated by the fact that the photon appears as a Nambu-Goldstone boson from this point of view. Remarkably, this also gives rise to a set of Buscher-like duality rules  \cite{Chatzistavrakidis:2020kpx}, which acquire a suggestive form when written in terms of the complexified coupling. Apart from discussing all these features, in this paper we make a first attempt to interpret generalized global symmetries as generalized isometries of the configuration space, at least in Abelian 1-form theories. Moreover, all these can be extended to higher degree differential forms, with even ones behaving like 2D scalars and odd ones like 4D 1-forms, cf. \cite{Julia:1980gr,Julia:1997cy,Deser:1997mz}. 

The rest of the paper is organized as follows. In Section \ref{sec2}, we review and revisit several aspects of T-duality in 2D scalar field theories, mainly highlighting the features that are usually left out in similar discussions. Section \ref{sec21} is a pedagogical introduction to the basics of the topic with a bird's eye view on the single compact scalar, its T-duality and 't Hooft anomaly between momentum and winding global symmetries. In Section \ref{sec22} we go beyond this stringy model and discuss the same aspects in a higher-derivative single scalar theory, the Heisenberg pion fireball model. In Section \ref{sec23} we explore in some detail the global symmetries of theories with multiple scalar fields and target space couplings that depend both on the fields and their derivatives. We argue that the correct geometrical formalism to account for this is jet bundles and jet prolongations. We then show that global symmetries correspond to generalized isometries in jet space. In Section \ref{sec24}, we briefly comment on the duality rules and the 't Hooft anomaly in multiscalar theories. Section \ref{sec3} establishes the corresponding results for Abelian, linear and nonlinear gauge theories in four dimensions. Sections \ref{sec31} and \ref{sec32} contain the general discussion on theories with electric/magnetic duality rotations, which is then applied to three illustrating examples in Section \ref{sec33}: the Maxwell, Born-Infeld and ModMax theories. In Section \ref{sec34}, we comment on the target space geometry of multiple Abelian 1-form gauge theories and use the jet space formalism to propose that their global symmetries also correspond to a notion of generalized Killing isometries in configuration space. Finally, Section \ref{sec4} contains a brief discussion on higher degree differential forms and mixed symmetry tensor fields and our general conclusions.   

\section{Features of duality in 2D scalar field theories}
\label{sec2}

\subsection{T-duality and the 't Hooft anomaly: basics}\label{sec21}

Our purpose in this section is to review some basic features of T-duality in 2D scalar field theories, namely nonlinear sigma models. Our main focus will be to highlight some of the features that are usually not covered in other treatments of the subject, such as the mixed 't Hooft anomaly between the momentum and winding symmetries and the T-duality of models whose couplings are not only functions of the scalar fields but of their derivatives as well. 

It is instructive to begin with the simplest and most pedagogical case of a single compact scalar field $X(\sigma^{\m})$, where $\sigma^{\m}, \mu=0,1$ are 2D spacetime (worldsheet) coordinates and the target space is a circle of radius $R$. In other words we consider a map $X:\S_{2}\to S^{1}$ and denoting the compact circle coordinate by $x \sim x+2\pi R$, the scalar field is the pullback $X(\s)=X^{\ast}(x)$. The simplest field theory in this setting is a kinetic term for $X$, the sigma model
\be \label{singlescalarlinearaction}
S[X]=-\frac 12 \int R^{2}\sqrt{|\gamma|} \gamma^{\m\n}\partial_{\m}X\partial_{\n}X\dd^2\s=-\frac 12 \int_{\S_{2}}R^2\,\dd X\w\ast\,\dd X\,,
\ee 
where we work in units where the Regge slope parameter is $\a'=1/2\pi$. Moreover, we have employed a differential form notation where the 2D metric $\gamma_{\m\n}$ and determinant $\g$ are packaged in the Hodge star operator $\ast$ and the associated volume form. Let us denote $F=\dd X$, essentially the field strength of the scalar field. At this stage it is useful to clarify that the symbol $F$ will be used in two different ways below. In accord with standard nomenclature, we refer to $F=\dd X$ as the \emph{constrained} 1-form or field strength, otherwise (i.e. when $F$ is not given by definition as $\dd X$, but eventually becomes $\dd X$ dynamically) we refer to it as the \emph{unconstrained} 1-form. 

The field equations obtained from the action functional \eqref{singlescalarlinearaction} accompanied by the Bianchi identity for the field strength $F$ read as 
\be \label{singlescalarlineareoms}
\dd\ast F=0 \quad \text{and}\quad \dd F=0\,.
\ee     
The theory under consideration has two global symmetries, the momentum and winding ones, with ordinary vector currents 
\be \label{singlescalarlinearcurrents}
J_{\text{mom}}= R^2 F \quad \text{and} \quad {J}_{\text{win}}= \ast F\,.
\ee 
Using the field equations \eqref{singlescalarlineareoms} it is evident that they are both conserved, $\dd \ast J_{\text{mom}}=0=\dd \ast{J}_{\text{win}}$. These two currents may be used to introduce background fields in the theory in the usual way of adding to the action terms of the form $A^{\m}J_{\m}$ or $A\w\ast J$ in differential form notation. In the present case the background fields are Abelian 1-forms and we denote them by $A$ and $\widehat{A}$, which will appear in the theory through $A\w\ast J_{\text{mom}}$ and $\widehat{A}\w\ast {J}_{\text{win}}$. These fields transform under background gauge transformations as 
\be \label{gauge trafos}
\d A=\dd \e \quad \text{and} \quad \d \widehat{A}=\dd \widehat{\e}\,,
\ee 
where $\e$ and $\widehat{\e}$ are the corresponding background gauge transformation parameters. Let us emphasize that being background fields, $A$ and $\widehat{A}$ are non-dynamical and they can be switched on and off at will without spoiling the theory and its symmetries. Nevertheless, introducing them in the theory by means of the terms $A\w\ast J_{\text{mom}}$ and $\widehat{A}\w\ast {J}_{\text{win}}$ requires a suitable counter term in order to preserve invariance under the transformation of $A$. Indeed, the dynamical field $X$ transforms as 
\be 
\d X=\e
\ee   
and is inert under the ``winding'' transformation given by $\widehat{\e}$, therefore the invariant 1-form is $\dd X-A$\,. This dictates that under suitable rearrangements of terms the action \eqref{singlescalarlinearaction} extended by background fields has the form 
\be \label{singlescalarlinearactionbackground}
S[X,A,\widehat{A}]=-\frac 12\int_{\S_2}R^2(\dd X-A)\w\ast\,(\dd X-A)+\int_{\S_2}\widehat{A}\w \dd X\,.
\ee   
To avoid confusion, we stress once more that this is not a gauged action and the background fields may be switched off without any issues. In particular, one should not think of $\widehat{A}$ here as a Lagrange multiplier. 

One may ask whether $A$ and $\widehat{A}$ could now be promoted to dynamical fields,{\footnote{One should note that even trying to do so the resulting theory would be topological, since we are in 2 dimensions. Here we use this terminology to relate this discussion to theories with $p$-forms. More precisely, the question here is whether we can promote the c-number background fields to operator-valued ones.}} in other words whether both global symmetries can be gauged. For this to be possible one should make sure that no 't Hooft anomaly is present in the theory. It is a known fact though that such an anomaly exists and prevents the gauging of both global symmetries simultaneously. This is easily seen by examining how the  action \eqref{singlescalarlinearactionbackground} changes under background gauge transformations. The result, discarding boundary terms, is 
\be \label{singlescalarlinearshift}
S[X+\e,A+\dd\e,\widehat{A}+\dd\widehat{\e}\,]=S[X,A,\widehat{A}\,]+\int_{\S_2}\e \, \dd \widehat{A}\,,
\ee 
where in the second term we took the liberty of integrating by parts, ignoring global issues that could however be treated in a proper way \cite{Hori:2000kt}, a topic that we shall not pursue further here. The main observation is that the action with background fields is not invariant but instead it includes shifts under the background gauge transformation of $A$.{\footnote{This is not at all similar to the case of Chern-Simons theory or the Wess-Zumino-Witten model, where the action is not invariant but the partition function is, under the assumption of integer coupling. In the present case the background field dependent partition function is not invariant too.}} Unlike the case of ABJ anomalies, these are c-number shifts and the anomaly goes away as long as the background fields are switched off. This is typical of 't Hooft anomalies, which represent an obstruction to gauging a global symmetry rather than a sickness in the theory. This is precisely what happens in the present case, as mentioned in footnote 11 of Ref. \cite{Gaiotto:2014kfa}. There exists a mixed 't Hooft anomaly between the momentum and winding global symmetries that prevents them from being gauged simultaneously. As discussed in \cite{Cordova:2018cvg}, one can present the anomalous term in multiple ways upon adding counter terms, however it can never be removed and persists in all descriptions of the theory, the latter being a manifestation of 't Hooft anomaly matching. In a more symmetric way, the shift in \eqref{singlescalarlinearshift} may be obtained via inflow from the 4-form anomaly polynomial 
\be \label{anomaly polynomial I4}
{\cal I}_4=\int_{\S_{4}}\dd A\w \dd\widehat{A}
\ee  
using the descent equations. 

The above discussion can now be directly related to the gauging of global symmetries in 2D sigma models. In this simple setting the situation is very clear, since the sigma model \eqref{singlescalarlinearaction} has a global (momentum) symmetry $\d X=\e$ with a rigid parameter $\e$. Gauging this symmetry means that $\e$ is promoted to a $\s$-dependent parameter which according to the basic principles of gauge theory means that we must introduce gauge field 1-forms ${\cal A}$ that transform as $\d {\cal A}=\dd \e$ and give rise to the covariant worldsheet differentials 
\be \mathrm{D} X=\dd X-{\cal A}\,. \ee 
Employing minimal coupling, the gauged action functional is 
\be 
S_{\text{gauged}}[X,{\cal A}]=-\frac 12 \int_{\S_{2}}R^2\, \mathrm{D}X\w\ast\,\mathrm{D}X\,.
\ee 
In this action, ${\cal A}$ is a true gauge field and not a background, which explains the difference in notation. One may now immediately see that 
\be 
S_{\text{gauged}}[X,{\cal A}]=S[X,A,\widehat{A}]|_{A\to{\cal A},\widehat{A}\to 0}\,.
\ee 
In words, the gauged action is obtained from the ungauged one coupled to background fields for all global symmetries after one of the background fields is promoted to a genuine gauge field and the other is switched off. This is the essence of the 't Hooft anomaly, which suggests that only one of the background fields can be promoted to a dynamical one. 

Of course one may ask whether we could switch off $A$ instead of $\widehat{A}$ and gauge the other global symmetry. This is better seen in terms of dual variables, which is where T-duality comes into play. Moreover, T-duality can be better understood in terms of an action involving the unconstrained 1-from $F$. To be fully general, we consider this ``parent'' action already coupled to background fields, 
\be \label{singlescalarlinearparent}
{\cal S}[F,\widehat{X},A,\widehat{A}\,]=-\frac 12 \int_{\S_{2}}R^{2}(F-A)\w\ast \,(F-A)+\int_{\S_2}F\w (\dd \widehat{X}-\widehat{A})\,,
\ee  
where $\widehat{X}$ is a Lagrange multiplier which will turn out to be the dual scalar field under T-duality. The action ${\cal S}$ is obtained from \eqref{singlescalarlinearactionbackground} upon replacing the constrained 1-form $F=\dd X$ by an unconstrained one and adding the $\widehat{X}$-dependent term $F\w\dd\widehat{X}$. Moreover, background gauge invariance dictates that the new field should transform according to 
\be \label{shift}
\d\widehat{X}=\widehat{\e}\,,
\ee 
a sign of the winding global symmetry as we discuss shortly. 
One can immediately see that variation with respect to the Lagrange multiplier $\widehat{X}$ leads to the Euler-Lagrange equation $\dd F=0$, which may be locally solved by the constrained $F=\dd X$ due to the Poincar\'e lemma. In that case the field $\widehat{X}$ is eliminated or integrated out from the theory and one returns to the original action \eqref{singlescalarlinearactionbackground}, or \eqref{singlescalarlinearaction} when the background fields are switched off. On the other hand, we can now instead eliminate the unconstrained field $F$ through its field equation. The latter is 
\be \label{singlescalarlineardualityrelation}
\ast(F-A)=\frac 1{R^2}(\dd\widehat{X}-\widehat{A})\,.
\ee     
This is substituted back in the action \eqref{singlescalarlinearparent} to give 
\be 
\label{singlescalarlinearbackgrounddual}
{S}[\widehat{X},A,\widehat{A}\,]=-\frac 12 \int_{\S_{2}}\frac 1{R^{2}}(\dd\widehat{X}-\widehat{A})\w\ast (\dd\widehat{X}-\widehat{A})+\int_{\S_2}A\w \dd \widehat{X}-\int_{\S_{2}}A\w\widehat{A}\,.
\ee  
This is then the action for the dual scalar field $\widehat{X}$ that transforms with shifts under the winding symmetry of the theory. One can now make the following three observations. First, switching off the background fields we immediately see that the dual scalar propagates on a target $S^{1}$ with radius $1/R$ (recall that we work in units where $\a'$ is 1/2$\pi$; from dimensional analysis the dual radius is actually $\a'/R$). Second, in this dual formulation of the theory the momentum and winding currents are exchanged. Indeed  the duality relation \eqref{singlescalarlineardualityrelation} taken with the constrained $F=\dd X$ says that the Bianchi identity and wave equation of the original theory are mapped to the wave equation and the Bianchi identity respectively in the dual theory, as in \cite{Duff:1989tf}. Third, the dual action contains a term $A\w\widehat{A}$ that couples the two background fields, unlike the action \eqref{singlescalarlinearactionbackground}. The reason for this can be traced in the matching of the 't Hooft anomaly, since 
\be 
S[\widehat{X}+\widehat{\e},A+\dd \e,\widehat{A}+\dd \widehat{\e}\,]=S[\widehat{X},A,\widehat{A}\,]+\int_{\S_2}\e\, \dd\widehat{A}\,.
\ee  
In other words, we observe that 
\be 
S[{X}+{\e},A+\dd \e,\widehat{A}+\dd \widehat{\e}\,]-S[{X},A,\widehat{A}\,]=S[\widehat{X}+\widehat{\e},A+\dd \e,\widehat{A}+\dd \widehat{\e}\,]-S[\widehat{X},A,\widehat{A}\,]\,,
\ee 
and the 't Hooft anomaly is precisely matched in both T-dual formulations of the theory. 

Summarizing, within the simple setting of a 2D compact scalar with target space being a circle we have discussed the global momentum and winding symmetries, their 't Hooft anomaly and the T-duality that exchanges the two symmetries along with the inversion of the radius of the target circle. This is the extension with background fields of the usual picture of equivalence between sigma models by means of a parent action functional found in string theory textbooks and can be depicted as 
\be\label{chain}
S[X]\overset{\cancel{A,\widehat{A}}}\longleftarrow S[X,A,\widehat{A}\,] \overset{\cancel{\widehat{X}}}\longleftarrow {\cal S}[F,\widehat{X},A,\widehat{A}\,] \overset{\cancel{F}}\longrightarrow S[\widehat{X},A,\widehat{A}\,]\overset{\cancel{A,\widehat{A}}}\longrightarrow S[\widehat{X}]\,.
\ee 
Here ${\cal S}$ is the parent action coupled to background fields, the inner arrows lead to the two dual theories coupled to background fields once by integrating out the field $\widehat{X}$ (left) and once the unconstrained field $F$ (right), and the outer arrows mean that the background fields are switched off and lead to the two dual second order theories in terms of $X$ and $\widehat{X}$ respectively. As we will see below, all the essential features of this picture and the relation of the duality to the 't Hooft anomaly are reflected in various generalizations of this basic model.   

\subsection{T-duality beyond strings: Heisenberg's pion fireball model}
\label{sec22}

The discussion in Section \ref{sec21} can be generalized to 2D higher-derivative single scalar theories in a straightforward way. We present this through the example of
Heisenberg's pion fireball model \cite{Heisenberg:1952zz}, which is based on the Dirac-Born-Infeld type action functional
\be \label{heisenberg}
S_H[X]=\int\ell^2\left(-1+\sqrt{1-\frac{R^2}{\ell^2}\,(\partial_\mu X\partial^\mu X+m^2X^2)}\right)\sqrt{|\g|}\,\dd^2\s\,,
\ee
where $\ell$ is a length parameter.  Here we are interested in the massless case, $m=0$.
Then, in the limit $\ell\to \infty$ one obtains the linear scalar theory \eqref{singlescalarlinearaction} studied in the previous section. Hence we set $R=1$ for simplicity. 
The field equation obtained by variation of this action with respect to $X$ and the Bianchi identity have the form
\be 
\ast\dd\ast\left(\frac{F}{\sqrt{1-\frac{1}{\ell^2}\,F_\mu F^\mu}}\right)=0\quad \text{and}\quad \dd F=0\,,
\ee
in terms of the constrained field strength $F=\dd X$. This theory has the same momentum and winding zero-form global symmetries as the linear theory \eqref{singlescalarlinearaction}, with corresponding vector currents 
\be \label{currents}
J_{\text{mom}}=\frac{F}{\sqrt{1-\frac{1}{\ell^2}\,F_\mu F^\mu}}\quad\text{and}\quad J_{\text{win}}=\ast F\,.
\ee
The first is conserved ($\dd\ast J_{\text{mom}}=0$) by virtue of the field equation, while the second is conserved ($\dd\ast J_{\text{win}}=0$) due to the Bianchi identity. 

Just like before, one can couple background fields $A$ and $\widehat{A}$ to the Heisenberg model. These transform under the background gauge transformations \eqref{gauge trafos} and the dynamical field $X$ transforms only under the first one through a shift. Since $X$ is inert under the winding transformation, the current $J_{\text{win}}$ can be coupled simply through the invariant (up to a total derivative) term $\widehat{A}\wedge\ast J_{\text{win}} $ and no additional counterterms are needed. On the other hand, the term $A\wedge\ast J_{\text{mom}}$ is not invariant on its own. Thus, one has to add an infinite sequence of counterterms which are of higher order $\mathcal{O}(A^2)$ to ensure invariance. These are all encoded in the following action
\be \label{heisenberg action with bg fields}
S_H[X,A,\widehat{A}]=\int\ell^2\left(-1+\sqrt{1-\frac{1}{\ell^2}\,(F_\mu-A_\mu)(F^\mu-A^\mu)}\right)\sqrt{|\g|}\,\dd^2\s+\int_{\Sigma_2}\widehat{A}\wedge F\,,
\ee
in which the term linear in $A$ reads precisely as $A\wedge\ast J_{\text{mom}}$. To see this, one can first expand the square root in the denominator and rewrite the components of the current as
\be
J_{\text{mom}}^\mu=F^\mu\sum_{m=0}^\infty\frac{1}{\ell^{2m}}\binom{m-1/2}{m}F^{2m}=\ell^2F^\mu\sum_{n=1}^\infty \frac{1}{\ell^{2n}}\binom{n-3/2}{n-1}F^{2n-2}\,,
\ee
where $F^{2m}\equiv(F_\mu F^\mu)^m$, etc.
Subsequently, we can expand the first term in the action as
\begin{equation}\begin{split}
\ell^2\left(-1+\sqrt{1-\frac{1}{\ell^2}\,(F_\mu-A_\mu)(F^\mu-A^\mu)}\right)&=\ell^2\sum_{n=1}^\infty\frac{(-1)^n}{\ell^{2n}}\binom{1/2}{n}(F-A)^{2n}\\
&=\ell^2\sum_{n=1}^\infty\sum_{k=0}^{2n}\frac{(-1)^{n+k}}{\ell^{2n}}\binom{1/2}{n}\binom{2n}{k}F^{2n-k}A^k\,.
\end{split}
\end{equation}
The terms linear in $A_\mu$, i.e. the ones with $k=1$ in the series, are 
\be
\ell^2F^\mu\sum_{n=1}^\infty \frac{(-1)^{n+1}}{\ell^{2n}}2n\binom{1/2}{n} F^{2n-2}\,.
\ee
Finally, the identity $\binom{n-3/2}{n-1}=2n(-1)^{n+1}\binom{1/2}{n}$ reveals that 
\be \ell^2F^\mu\sum_{n=1}^\infty \frac{(-1)^{n+1}}{\ell^{2n}}2n\binom{1/2}{n} F^{2n-2}=A_\mu J_{\text{mom}}^\mu
\ee
and this concludes the proof of the statement.

As we saw earlier for the case of the string sigma model, the action is not invariant under background gauge transformations,  since 
\be 
S_H[X+\e,A+\dd\e,\widehat{A}+\dd\widehat\e\,]=S_H[X,A,\widehat{A}\,]+\int_{\S_2}\e\,\dd\widehat{A}\,,
\ee
up to boundary terms. The above shift may be obtained via inflow from the same $4$-form anomaly polynomial $\mathcal{I}_4$ given in \eqref{anomaly polynomial I4}. This reflects the fact that in both the string sigma model and in this particular higher-derivative theory the momentum and winding global symmetries are the same.
If one now wishes to gauge part of the global symmetry, using minimal coupling the gauged action reads as
\be 
S_{H,\,\text{gauged}}[X,\mathcal{A}]=\int \ell^2\left(-1+\sqrt{1-\frac{1}{\ell^2}\,\mathrm{D}_\mu X\mathrm{D}^\mu X}\right)\sqrt{|\g|}\,\dd^2\s\,.
\ee
 Once more the gauged action $S_{H,\,\text{gauged}}[X,\mathcal{A}]$ is obtained from $S_H[X,A,\widehat{A}\,]$ by promoting the background field $A$ to the dynamical gauge field $\mathcal{A}$ and switching off the background field $\widehat{A}$. 

To uncover T-duality in this case, let us consider the parent action
\be 
\mathcal{S}_H[F,\widehat{X},A,\widehat{A}\,]=\int\ell^2\left(-1+\sqrt{1-\frac{1}{\ell^2}\,(F_\mu-A_\mu)(F^\mu-A^\mu)}\right)\sqrt{|\g|}\,\dd^2\s+\int_{\Sigma_2}F\wedge(\dd\widehat{X}-\widehat{A})\,,
\ee
which is already coupled to the background fields $A$ and $\widehat{A}$. The field $F$ is now an unconstrained $1$-form and $\widehat{X}$ is a Lagrange multiplier, which will be identified with the dual scalar field. The original second order theory \eqref{heisenberg action with bg fields} is obtained by varying the parent action with respect to the Lagrange multiplier $\widehat{X}$, as before. On the other hand, the equation of motion for $F$ is the duality relation 
\be \label{dr}
\frac{F^\mu-A^\mu}{\sqrt{1-\frac{1}{\ell^2}\,(F_\nu-A_\nu)(F^\nu-A^\nu)}}=\epsilon^{\m\n}(\partial_\n\widehat{X}-\widehat{A}_\n)\,.
\ee
Multiplying both sides of this relation with $(F_\m-A_\m)$ and $\e_{\m\k}(\partial^\k\widehat{X}-\widehat{A}^\k)$ respectively gives the two equations
\be \begin{split}
\frac{\alpha}{\sqrt{1-\frac{1}{\ell^2}\alpha}}&=\e^{\m\n}(F_\m-A_\m)(\partial_\n\widehat{X}-\widehat{A}_\n)\,,\\
\widehat{\alpha}\sqrt{1-\frac{1}{\ell^2}\alpha}&=-\e^{\m\n}(F_\m-A_\m)(\partial_\n\widehat{X}-\widehat{A}_\n)\,,
\end{split}
\ee
where we have introduced the shorthand notation
\be 
\alpha:=(F_\mu-A_\mu)(F^\mu-A^\mu)\quad\text{and}\quad \widehat{\alpha}:=(\partial_\mu\widehat X-\widehat A_\mu)(\partial^\mu\widehat{X}-\widehat A^\mu)
\ee
for brevity. These equations imply that 
\be 
\widehat{\alpha}=-\frac{\alpha}{1-\frac{1}{\ell^2}\alpha}\qquad\Longleftrightarrow\qquad\alpha=-\frac{\widehat{\alpha}}{1-\frac{1}{\ell^2}\widehat{\alpha}}\,.
\ee
Using this final equation, as well as the duality relation \eqref{dr} from which it originates, one can obtain the dual theory from the parent action by direct substitution. The result is
\be 
S_H[\widehat{X},A,\widehat{A}]=\int\ell^2\left(-1+\sqrt{1-\frac{1}{\ell^2}\,(\partial_\mu\widehat{X}-\widehat A_\mu)(\partial^\mu\widehat{X}-\widehat A^\mu)}\right)\sqrt{|\g|}\,\dd^2\s+\int_{\Sigma_2}A\wedge(\dd\widehat X-\widehat{A})\,.
\ee
We observe once again that the dual action contains the term 
$-\int_{\S_2} A\wedge \widehat{A}$,
which ensures the matching of the corresponding 't Hooft anomaly. Finally, it is clear that the equivalence reflected in \eqref{chain} holds for the Heisenberg model too, with the same radius inversion as in the string model.

\subsection{Multiple fields, target space geometry and jets}\label{sec23}

A further natural generalization of the models discussed so far regards theories with multiple fields and  richer target space isometries. In that case the sigma model map $X:\S_2\to M$ has a target manifold $M$ replacing the circle $S^{1}$. The components of the map $X$ are scalar fields $X^{i}$ with $i=1,\dots,\text{dim}\,M$, not all of which need to be necessarily compact. However, when some of them are compact, dualization may be performed along these directions.\footnote{Compactness, or periodicity, of the scalar fields guarantees the existence of momentum and winding global symmetries and moreover that the spectrum of the theory after quantization is invariant under dualization and exchange of momentum and winding modes.}
We consider a nonlinear, possibly higher-derivative sigma model for the scalar fields $X^i$ given by the action functional
\begin{align}\label{eq:action-mult-scalar}
S[X] =- \frac{1}{2}\int_{\Sigma_2} \bigg(G_{ij}(X,\dd X)\, \dd X^i \wedge \ast\, \dd X^j + B_{ij}(X,\dd X)\, \dd X^i \wedge \dd X^j\bigg),
\end{align}
where $G_{ij}$ and $B_{ij}$ are background fields which play the role of couplings in the 2D theory. In usual cases, such as for closed strings in background fields, these couplings are functions of $\sigma^{\m}$ only through the scalar fields $X^{i}$. Here we consider an extension of this picture to theories where the background fields can also depend on $\s^{\m}$ via the worldsheet differentials (field strength) $\dd X^{i}$. We have already seen a realization of this when we discussed the Heisenberg pion fireball model.

First, it is important to identify under which conditions this action has some global symmetries. 
In the case where $G_{ij}$ and $B_{ij}$ are either constant or strictly $X$-dependent, the answer is well known, see for instance \cite{Giveon:1994fu,Giveon:1991jj} and references therein. Geometrically, $G_{ij}$ are the pull-back components of a Riemannian metric $G$ on $M$ and $B_{ij}$ the ones of a 2-form $B$, with field strength $H=\dd B$.  
Then the action is invariant under the global symmetry
 \be 
 \delta X^i = \rho^i_a(X)\, \epsilon^a\,,
 \ee 
  provided that the following invariance conditions on the background fields hold
\begin{align}
 \mathcal{L}_{\rho_a} G &= 0\,,\label{eq:Lie_bg_Field_G}\\[4pt]
  \mathcal{L}_{\rho_a} B &= \dd \beta_a \quad \Leftrightarrow \quad {\cal L}_{\rho_a}H=0\,.\label{eq:Lie_bg_Field_B}
\end{align}  
Here $\epsilon^a$ is the (rigid) parameter of the global symmetry and $\rho^i_a$ are the components of a Lie algebra valued vector field $\rho$ in the configuration space of fields, i.e. $\rho = \rho^i \partial_{i}$ and $\rho^i = \rho^i_a t^a$ with $t^a$ generators of the global symmetry.{\footnote{An extension of these statements to (almost) Lie algebroids and singular foliations instead of Lie algebras was considered in \cite{Chatzistavrakidis:2016jfz}, but this lies beyond the scope of the present paper.}} Furthermore, $\beta_a$ is an arbitrary 1-form on $M$. Recall that the Lie derivative of e.g. the metric $G$ in a local coordinate system reads as 
\be 
({\cal L}_{\rho_a}G)_{ij}=\rho_a^{k}\partial_kG_{ij}+G_{ik}\partial_{j}\rho_{a}^{k}+G_{kj}\partial_{i}\rho_a^{k}\,.
\ee 
This tells us that the global symmetries of the theory are isometries of the target space geometry specified by $G$ and $H$ (a generalized metric in the language of \cite{Severa:2019ddq}), and $\rho_a$ is thus a host of Killing vector fields. Another way to think about this is to rewrite \eqref{eq:Lie_bg_Field_B} as 
\be 
\iota_{\rho_a}H=\dd\a_{a}\,,
\ee 
where $\a_{a}=\beta_a-\iota_{\rho_a}B$. Then one may think of $\rho_a+\a_a$ as a section of the generalized tangent bundle $TM\oplus T^{\ast}M$ and the two invariance conditions can be combined into 
\be 
[\rho+\a,E]=0\,,
\ee 
where $E=G+B$ and the bracket is the one on an exact Courant algebroid. In other words, this represents a generalized flow for the tensor field $E$, see \cite{Streets:2013uda,Severa:2016lwc} for more details.  
The single compact scalar is obviously a special case of this with the isometry being the one on the circle $S^{1}$, in which case the invariance conditions are trivially satisfied. 
In the following, we revisit these well-known statements for the apparently more general case of background couplings $G_{ij}, B_{ij}$ that depend both on the fields $X^{i}$ and their field strength $\dd X^{i}$, simultaneously generalizing the closed string on a circle and the Heisenberg model. In this case, one must treat $X^i$ and $\dd X^i$ as independent fields. There is a natural mathematical framework that allows us to deal with the variational principle of the action where fields and their higher derivatives are independent. It goes under the name of ``variational bicomplex'' \cite{Anderson:1992} which is based on an extension of the configuration space to the jet space. Below we provide a basic toolkit for our purposes.

Let us consider a smooth fibre bundle  $(E,\pi,\Sigma)$ with base $\Sigma$, fibre $M$ and projection $\pi: E \to \Sigma$, and a point $p\in \Sigma$. Two local sections $X,Y\in \Gamma_p(\pi)$ are called $1$-equivalent\footnote{These two conditions are indeed the first two terms in the Taylor expansion of $X,Y$ around $p$. One can similarly define $n$-equivalent or $\infty$-equivalent sections.} at $p$ if $X(p) = Y(p)$ and in an adapted coordinate system $(\sigma^{\mu},x^i)$ of the bundle around $X(p)$
\begin{equation}
\left.\frac{\partial X^i}{\partial \sigma^{\mu}}\right\vert_p = \left.\frac{\partial Y^i}{\partial \sigma^{\mu}}\right\vert_p,
\end{equation}
for $i=1,...,\textrm{dim}(M)$ and $\mu = 1,...,\textrm{dim}(\Sigma)$ and $X^i = X^*(x^i)$. This equivalence class is called the 1-jet of $X$ at $p$ and is denoted by $j^1_p X$. The first jet manifold of $\pi$ is the set 
\begin{align}
J^1\pi:\left\lbrace j^1_p X: p\in \Sigma, X \in \Gamma_p(\pi)\right\rbrace.
\end{align}
The functions $\pi_1$ and $\pi_{1,0}$ defined by 
\begin{align}
\pi_1 &: J^1 \pi\rightarrow \Sigma,\qquad \text{s.t.} \qquad \pi_1(j^1_p X) = p\,,\\
\pi_{1,0} &: J^1 \pi\rightarrow E,\qquad \text{s.t.} \qquad  \pi_{1,0}(j^1_p X) = X(p)\,
\end{align}
are called the source and target projections respectively. 
Now let $(U,u)$ be an adapted coordinate system on $E$, where $u=(\sigma^\mu,x^i)$ and $U$ is an open neighbourhood around $X(p)$. The induced coordinate system $(U',u')$ on $J^1 \pi$ is defined by
\begin{align}
U' &=\left\lbrace j^1_p X: X(p)\in U\right\rbrace,\\
u' &=\left(\sigma^\mu, x^i, x^i_{\mu}\right),
\end{align}
where the new functions $x^i_{\mu}:U'\to \mathbb{R}$ are specified by
\begin{equation}
x^i_{\mu} =\left.\frac{\partial X^i}{\partial \sigma^\mu}\right\rvert_p = \left.\partial_{\mu} X^i \right\rvert_p \equiv \left. X^i_\mu\right\rvert_p,
\end{equation}
and are known as derivative coordinates. It can be shown that the triples $(J^1 \pi, \pi_1, \Sigma)$ and $(J^1 \pi,\pi_{1,0},E)$ are also fibred manifolds. These steps can be repeated $n$ times in order to form an $n$-jet bundle.

In the same manner that $(\frac{\partial}{\partial \sigma^{\mu}},\frac{\partial}{\partial x^i})$ form a complete basis for a fibre bundle, one can generalize this to an $n$-jet space with the basis
\begin{equation}
(\frac{\partial}{\partial \sigma^{\mu}},\frac{\partial}{\partial x^i},\frac{\partial}{\partial x^i_{\mu_1}},\frac{\partial}{\partial x^i_{\mu_1\mu_2}},...,\frac{\partial}{\partial x^i_{\mu_1\mu_2...\mu_n}}).
\end{equation}
A general vector field in this basis can be written as
\begin{equation}\label{eq:gen-vec-field}
 V^{(n)} = V^{\mu} \frac{\partial}{\partial \sigma^{\mu}} + V^i \frac{\partial}{\partial x^i} + ... + V^i_{\mu_1...\mu_n} \frac{\partial}{\partial x^i_{\mu_1\mu_2...\mu_n}}\,.
\end{equation}
At this point its components are completely arbitrary. For $n=1$, the vector field $V^{(1)} \in \mathfrak{X}(J^1 \pi)$ is projectable if $V^\mu$, $V^i$ are smooth functions on $U$ and $V^i_{\mu}$ is a smooth function on $U'\equiv J^1(U)$. However, there are two ways of extending the coordinate space to the one on the jet-bundle which restricts the general form of the vector field, as we briefly explain below.

First, considering the bundle $(T^\ast E, \pi^\ast_E, E)$ the total space $\pi_{1,0}^\ast(T^\ast E)$ can be interpreted as a submanifold of $T^\ast J^1 \pi$. This allows us to define differential forms on $J^1 \pi$. A contact 1-form $\omega^i_{\mu_1...\mu_k}$ can be written in local coordinates as $\omega^i_{\mu_1...\mu_k} = dx^i_{\mu_1...\mu_k} - x^i_{\mu_1...\mu_k\nu} d\sigma^{\nu}$. Then a \emph{total vector field} $V^{(n)}$ on an $n$-jet space is a vector field of the form \eqref{eq:gen-vec-field} such that for a contact form $\omega^i_{\mu_1...\mu_k}$, it satisfies $\omega(V^{(n)}) =0$. This restricts the form of $V^{(n)}$ to
\begin{equation}
V^{(n)} = V^{\mu} \frac{\partial}{\partial \sigma^{\mu}} + V^{\mu} x^i_{\mu} \frac{\partial}{\partial x^i} + ... + V^{\mu} x^i_{\mu\mu_1...\mu_n} \frac{\partial}{\partial x^i_{\mu_1\mu_2...\mu_n}}\,,
\end{equation} 
with the condition that $x^i_{\mu_1\mu_2...\mu_k\nu} = - x^i_{\nu\mu_1...\mu_k}$. More precisely, the total vector fields are the sections of bundle of holonomic tangent vectors which is a pullback of the bundle of vertical tangent vectors in the sense that they are in the kernel of the map $\pi_\ast: TE \to T\Sigma$, and the contact forms are the sections of bundle of contact cotangent vectors which is a pullback of bundle of the horizontal cotangent vectors. These contact forms are annihilated by total vector fields. In the following, however, we will not use this extension of vector space but rather the prolongation approach described below.  

Another way of extending the basis to a complete basis in the jet space is through the prolongation of the vector space spanned by $(\frac{\partial}{\partial \sigma^{\mu}},\frac{\partial}{\partial x^i})$. Corresponding to each local section of the bundle $\pi$, there is a uniquely determined local section of the bundle $\pi_1$. This new section is called first prolongation and its coordinate representation is obtained by appending to the coordinates of the original section the derivatives of those coordinates. If $(E,\pi,\Sigma)$ is a bundle, $W \subset \Sigma$ is an open submanifold and $X \in \Gamma_W(\pi)$, then the first prolongation of $X$ is the section $j^1 X \in \Gamma_W(\pi_1)$ defined by $j^1 X(p) = j^1_p X$ for $p\in W$. Therefore, the coordinate representation of $j^1 X$ is $(X^i,\frac{\partial X^i}{\partial \sigma^\mu})$. A general local section $\psi \in \Gamma_W(\pi_1)$ will have coordinates $(\psi^i, \psi^i_\mu)$ where $\psi^i_\mu$ is completely independent of $\psi^i$. Let us consider a vector field $V \in \mathfrak{X}(E)$ with the coordinate representation
\be
V = V^\mu \frac{\partial}{\partial \sigma^\mu} + V^i \frac{\partial}{\partial x^i}\,.
\ee
Then the prolongation of $V$ is $V^{(1)} \in \mathfrak{X}(J^1\pi)$ and can be represented in coordinates as \cite{Olver:1986,Saunders:1989}
\be\label{eq:vec-prolong}
V^{(1)} = V^\mu \frac{\partial}{\partial \sigma^\mu} + V^i \frac{\partial}{\partial x^i} +\left(\frac{d V^i}{d\sigma^\mu} - x^i_{\nu} \frac{d V^\nu}{d \sigma^\mu}\right)\frac{\partial}{\partial x^i_{\mu}}\,,
\ee
where $\frac{d}{d\sigma^\mu}$ is a total derivative defined as
\begin{equation}
\frac{d}{d\sigma^\mu} = \frac{\partial}{\partial\sigma^\mu} + x^i_{\mu}\frac{\partial}{\partial x^i}\,.
\end{equation}
A local basis for the full exterior algebra $\Omega(J^n \pi)$ can be given by the differential forms 
\begin{equation}
\dd\sigma^\mu, \dd x^i, \dd x^i_{\mu},..., \dd x^i_{\mu_1...\mu_n}.
\end{equation}

After this short review of the jet-bundle, we are ready to go back to the original problem of what is the global symmetry of the action \eqref{eq:action-mult-scalar} and how it is related to isometries of the target space geometry. The independent fields of the model are $X^i$ and $\dd X^i$, therefore we are extending the fibre bundle to a $1$-jet bundle. In the following we use $X^i_{\mu}$ instead of $\dd X^i$, where we consider $\dd X^i = X^i_\mu\, \dd\sigma^\mu$ and $\dd\sigma^\mu$ contributes to the volume-form on the worldsheet. Moreover, both $X^i$ and $X^i_\mu$ are functions of $\sigma^{\mu}$. This means that we are considering a jet space defined via the basis\footnote{More precisely, the basis is given by $\left\lbrace\partial_{x^i},\partial_{x^i_{\mu}}\right\rbrace$ and since $X^i = X^*(x^i) $ with the abuse of notation we use $\left\lbrace\partial_{X^i},\partial_{X^i_\mu}\right\rbrace$ as the basis.}$\left\lbrace\partial_{X^i},\partial_{X^i_\mu}\right\rbrace$. Consider now the following vector field defined using the prolongation \eqref{eq:vec-prolong} written in this basis,
\begin{equation}
V^{(1)} = \Lambda^{\mu} \frac{\partial}{\partial \sigma^\mu} + \rho^i \frac{\partial}{\partial X^i} +\xi^i_{\mu} \frac{\partial}{\partial X^i_{\mu}}\,,
\end{equation}
with $\Lambda^\mu,\rho^i,\xi^i_\mu$ to be Lie algebra valued functions of $\sigma^\mu, X^i$ in general. 
The transformation of the fields is
\begin{align}
\delta X^i &= \rho^i_a \epsilon^a\,,\\
\delta X^i_{\mu} &= \xi^i_{\mu a} \epsilon^a = \left(\frac{d \rho^i_a}{d\sigma^\mu} - X^i_{\nu} \frac{d \Lambda^\nu_a}{d \sigma^\mu}\right)\epsilon^a\nn\\
&\,~~~~~~~~~~=\left[\left(\partial_{\mu}\rho^i_a + X^j_\mu \frac{\partial \rho^i_a}{\partial X^j}\right) - X^i_{\nu}\left(\partial_{\mu}\Lambda^{\nu}_a + X^k_{\mu} \frac{\partial \Lambda^{\nu}_a}{\partial X^k}\right)\right]\epsilon^a\,.
\end{align}
In order to make contact with the problem of symmetries of the action \eqref{eq:action-mult-scalar}, we restrict to the case where $\rho^i$, $\xi^i_\mu$ do not depend explicitly on $\sigma^\mu$, therefore $\partial_{\mu} \rho^i =0$. Moreover, we restrict to the prolongation where $\Lambda^\mu =0$. Then the candidate global symmetry becomes 
\begin{align}
\delta X^i &= \rho^i_a \epsilon^a\,,\label{eq:global_sym_mul_scalar_1}\\
\delta X^i_\mu &= \frac{\partial \rho^i_a}{\partial X^j} X^j_{\mu} \epsilon^a\,.\label{eq:global_sym_mul_scalar_2}
\end{align}
By direct calculation the action \eqref{eq:action-mult-scalar}
is invariant under this global  symmetry
if and only if the following generalized invariance conditions hold
\begin{align}
(\mathcal{L}_{\rho} G)_{ij} + \xi^k_{\mu}\, \frac{\partial G_{ij}}{\partial X^k_{\mu}} &= 0\,,\label{eq:example_cond_G1}\\
(\mathcal{L}_{\rho} B)_{ij} + \xi^k_{\mu}\, \frac{\partial B_{ij}}{\partial X^k_{\mu}} &= \partial_{[i}\beta_{j]}\,,\label{eq:example_cond_B1}
\end{align}
in terms of the usual Lie derivative on the space with coordinate basis $\partial_{X^i}$.

These conditions acquire a more natural geometric interpretation in the $1$-jet space, using the Lie derivative of a differentiable multilinear map $T$ of smooth sections of the $1$-jet tangent and cotangent spaces defined using $\pi_1$ and $\pi_{1,0}$ into $\mathbb{R}$, i.e. $T:\otimes^{q}\Gamma(T^\ast\Sigma)\times \otimes^{p}\Gamma(T\Sigma) \to \mathbb{R}$. If $\omega_1,...\omega_p \in \Omega(J^1 \pi)$ and $V_1,V_2,...,V_q \in \mathfrak{X}(J^1 \pi)$, then the Lie derivative along some vector field $V$ in the 1-jet space is given\footnote{We note that the Lie derivative of a contact form along a prolonged vector field preserves the contact ideal, i.e. it is itself a contact form.} as
\begin{align}
(\widehat{\mathcal{L}}_V T)(\omega_1,...,\omega_p,V_1,...,V_q) &= V(T(\omega_1,...,\omega_p,V_1,...,V_q))\nn\\
&- \sum_{i=1}^p  T(\omega_1,..., \widehat{\mathcal{L}}_V \omega_i,...,\omega_p,V_1,...,V_q)\nn\\
&- \sum_{i=1}^q  T(\omega_1,..., \omega_p, V_1,..., \widehat{\mathcal{L}}_V V_i,...,V_q)\,.\label{jetlie}
\end{align}
This formula also holds for non-projectable vector fields since the right hand side of this equation is still well-defined \cite{Saunders:1989}.
Using this definition and the discussion above, it becomes obvious that the 
invariance conditions \eqref{eq:example_cond_G1} and \eqref{eq:example_cond_B1} acquire the form 
\begin{align}
\widehat{\mathcal{L}}_V G &= 0\,,\\
\widehat{\mathcal{L}}_V B &= \dd\beta\,,
\end{align}
for $G$ and $B$ the ordinary Riemannian metric and Kalb-Ramond field on the target space $M$. 
This is the main result of the present section. It extends the well-known result of just $X$-dependent background fields to the more general case of $(X,\dd X)$-dependent ones, such as the Heisenberg model and generalizations thereof. Remarkably, the usual statement about the global symmetries being Killing vector isometries of the target space $M$ is promoted to Killing vector isometries of the 1-jet for the mapping space.   
As for the single compact scalar with the circle of radius $R$ as target space, which is the prototypical example satisfying the usual invariance conditions trivially, the massless Heisenberg model does so for these extended invariance conditions. The model itself may be written in  the form of a nonlinear sigma model
\be \label{heisenberg sigma model}
S_H[X]=-\frac{1}{2}\int_{\S_2}G(\dd X)\,\dd X\wedge \ast\, \dd X\,,\quad G(\dd X)=\sum_{n=1}^\infty \frac{(-1)^n}{\ell^{2n}}\binom{1/2}{n}(\partial_\mu X \partial^\mu X)^{n-1}\,,
\ee
which is simply a rewriting of \eqref{heisenberg}, obtained by Taylor-expanding the square root. In this example, there only exists a single scalar field and, thus, there is no coupling $B$.
In the above language, the vector $V$ in the $1$-jet space reads as 
\be V=\rho\,\frac{\partial}{\partial X}+\xi_\mu\frac{\partial}{\partial F_\mu}\,. 
\ee 
The coupling is now $G=G(F_\mu)$, with $\frac{\partial G}{\partial X}=0$ and the derivatives of $X$ are assumed to be independent variables, i.e. $\partial_\mu X\equiv F_\mu$. The generalized Lie derivative along this vector is now simply given by 
\bea \label{eq:Heisenberg_model}
\widehat{\mathcal{L}}_VG&=&2G\frac{\partial\rho}{\partial X} +\xi_{\mu} \frac{\partial G}{\partial F_{\mu}} \nn\\ &=&2\sum_{n=1}^\infty \frac{(-1)^n}{\ell^{2n}}\binom{1/2}{n}(F_\m F^\m)^{n-2}\left[\frac{\partial\rho}{\partial X}F_\n F^\n +(n-1)\,\xi_\nu F^\nu\right]\,.
\eea
This vanishes if and only if 
\be\label{eq:cond_Heisenberg_Lie}
\frac{\partial\rho}{\partial X}F_\n F^\n +(n-1)\,\xi_\nu F^\nu=0\,,\qquad \forall n=1,\dots,\infty\,.
\ee
Given the global symmetry \eqref{eq:global_sym_mul_scalar_2}, we have $\xi_\nu = \frac{\partial \rho}{\partial X} F_\nu$, from which one can simplify further \eqref{eq:cond_Heisenberg_Lie} to
\be
n\frac{\partial\rho}{\partial X}F_\n F^\n =0\,,\qquad \forall n=1,\dots,\infty\,. 
\ee
Obviously, the only solution to this condition is $\frac{\partial\rho}{\partial X}=0$ and $\rho$ can be scaled away such that the global shift symmetry is given by $\delta X = \epsilon$, as in Section \ref{sec22}. This is, however, a somewhat degenerate example because $G$ is only a function of derivative coordinates and does not reflect the full power of jet-space formulation. On one hand, that would be the case for effective theories that involve couplings which are functions of both fields and their derivatives. 
Possible candidates for such effective theories are the so-called generalized Galileons \cite{Deffayet:2013lga}.

On the other hand, the transformation of $X^i_{\mu}$ can be written in even more general way by i) considering two different symmetry generators, i.e. for example $\delta X^i_\mu = \xi^i_{\mu a}\epsilon^a + \xi^i_{\mu J}\lambda^J$ where $\lambda^J$ are parameters of another type of symmetry, ii) allowing $\Lambda^{\mu} \neq 0$ or $\partial_{\mu} \rho^i \neq 0$. A theory with  Galilean shift symmetry \cite{Nicolis:2008in}, or more general polynomial shift symmetries \cite{Griffin:2014bta}, is an example of this extended notion of transformation. A Galilean symmetry is described by $X^i_{\m} \to X^i_{\m}+ b^i_\mu $, where $b^{i}_{\m}$ are constants. In our notation, this can be written as
\begin{align}
\delta X^i &=\rho^i(\sigma^\mu,X^j),\nn\\
&=\hat{\rho}^i_a(X^j) \epsilon^a + \tilde{\rho}^{i\mu}_j(\sigma^\nu) b^j_\mu.
\end{align}
This extra term in the transformation of $X^i$ inspires new terms in the transformation of $X^i_\mu$ which would then lead to more stringent geometrical constraints on the target space geometry. We leave a complete discussion of these apparently more general theories for future work.

\subsection{T-duality and the 't Hooft anomaly: multiple fields}
\label{sec24}

Having identified the global symmetry of the multifield action \eqref{eq:action-mult-scalar}, we can now complete the discussion by first coupling spacetime background fields to it along the same lines as in sections \ref{sec21} and \ref{sec22}, and then stating the duality rules for the target space background fields. In this section we focus on the case where $G_{ij}, B_{ij}$ are only functions of the dynamical fields $X^i$. Therefore, we restrict to those transformations under which $X^i$ transform
as
$\delta X^i = \rho^i_a \epsilon^a$
and the background fields $A^a$ and $\widehat{A}_a$ transform as
\be
 \delta A^a = \dd\epsilon^a\quad \text{and} \quad \delta \widehat{A}_a = \dd\widehat{\epsilon}_a.
\ee
Here the momentum and winding parameters $\epsilon^a$ and $\widehat{\epsilon}^{\,a}$ are functions of the spacetime coordinates $\sigma^\mu$.
The action \eqref{eq:action-mult-scalar} can now be coupled to these background fields as
\begin{align}\label{eq:action-mult-scalar-bg}
S[X,A,\widehat{A}] = -\frac{1}{2}\int_{\Sigma_2} &\bigg(G_{ij}\, (\dd X^i - \rho^i_a A^a) \wedge \ast (\dd X^j -\rho^j_a A^a) +  B_{ij}\, (\dd X^i - \rho^i_a A^a) \wedge (\dd X^j -\rho^j_a A^a)\nn\\
&-\beta_{ai} (\dd X^i - \rho^i_b A^b) \wedge A^a - \delta^a_i \dd X^i \wedge \widehat{A}_a\bigg).
\end{align}
For vanishing background fields, the field equation for the scalars reads as 
\be 
 \dd\ast \dd X^i+\Gamma_{jk}^i\,\dd X^j\wedge\ast\dd X^k=\frac{1}{2}(G^{-1})^{il}H_{jkl}\,\dd X^j\wedge \dd X^k\,,
\ee
where $\G^{i}_{jk}$ are the connection coefficients obtained from the metric $G$ and $H_{ijk}$ are the components of the 3-form field strength of $B$. It is then straightforward to identify the two currents in this case, which read as
\begin{equation}
    J_a^{\text{mom}}=(\iota_{\rho_a} G)_j\,F^j+(\iota_{\rho_a} B-\beta_a)_j\ast F^j \quad \text{and} \quad J^a_{\text{win}}=\delta^a_i\ast F^i
\end{equation}
in terms of the constrained $F^{i}$. The conservation of these currents is a direct consequence of the field equation and the Bianchi identity respectively. The action \eqref{eq:action-mult-scalar-bg} contains the usual terms that couple background fields through these currents, improved by additional counter terms. 
With the already stated invariance conditions on the couplings $G$ and $B$, this action is invariant under the background gauge transformation only when $\widehat{A}_a$ is set to zero. 
The transformation of the full action \eqref{eq:action-mult-scalar-bg} (which involves the term $\ast J_{m}\wedge \widehat{A}_a$) 
under the background gauge symmetry is non-vanishing, specifically
\begin{align}
\delta S = \int_{\Sigma_2} \delta^a_i \rho^i_b \epsilon^b \dd\widehat{A}_a\,.
\end{align}
Using adapted coordinates along the isometry directions such that $\rho^{i}_{a}=\d^{i}_{a}$, this anomalous term can be obtained from an anomaly polynomial in $4$ dimensions
\begin{equation}
\mathcal{I}_4 =  \int_{\Sigma_4} \dd A^a \wedge \dd\widehat{A}_a\,.
\end{equation}
We thus observe that more possibilities arise in the multifield case when one wishes to gauge part of the full global symmetry. Apart from electric and magnetic gaugings, one may consider dyonic ones where some of the $A$ and $\widehat{A}$ fields are kept and promoted to true gauge fields, whereas their conjugates in the anomaly polynomial are set to zero.

The action \eqref{eq:action-mult-scalar-bg} can be now dualized along the isometric directions with the same procedure as in Sections \ref{sec21} and \ref{sec22}. As before, the dual theory will have the same 't Hooft anomaly with the original one but it will comprise different, T-dual couplings $\widehat{G}_{ij},\widehat{B}_{ij}$. As discussed in textbooks, see e.g. \cite{Blumenhagen:2013fgp}, the dual action is 
\begin{align}
\widehat{S}[\widehat{X}] = -\frac{1}{2}\int_{\Sigma_2} \bigg(\widehat{G}_{ij}(\widehat{X})\, \dd\widehat{X}^i \wedge \ast \,\dd\widehat{X}^j + \widehat{B}_{ij}(\widehat{X})\, \dd\widehat{X}^i \wedge \dd\widehat{X}^j\bigg)\,,
\end{align}
where $\widehat{X}^{i}=(X^{\a},\widehat{X}_{m})$ are dual dynamical fields, differing from the original ones $X^{i}=(X^{\a},X^{m})$ only in the directions $m=1,\dots d$ that undergo duality. The procedure establishes that the dual couplings are related through the Buscher rules, which in terms of  the generalized metric 
\be
E_{ij} = G_{ij} + B_{ij}\,,
\ee
 can be written as
\be
\widehat{E}_{ij} = E_{ij} - E_{im}E^{mn}E_{nj},\quad \widehat{E}_{im} = E_{in} E^{nm},\quad \widehat{E}_{mn} = E^{mn},
\ee
where $m,n$ run from 1 to the number $d$ of directions along which the dualization is performed. These are $O(d,d,\mathbb{R})$ transformations of the generalized metric. We note in passing that this set of rules holds true for all corresponding sigma models of differential $2p$-form fields in self-dual dimensions, e.g. 2-forms in 6 dimensions, with a kinetic term and a generalized theta term that corresponds to the antisymmetric coupling $B_{ij}$. For more details see \cite{Giveon:1994fu,Chatzistavrakidis:2020kpx}.

\section{Nonlinear theories of an Abelian 1-form \& duality}
\label{sec3}

\subsection{General setting}
\label{sec31}

Let us now move on to 4D nonlinear Abelian gauge theories admitting SO(2) electric-magnetic duality rotations and present a systematic procedure for their off-shell dualization. This essentially extends the results presented in \cite{Chatzistavrakidis:2020kpx} for linear theories and it reflects the general spirit of e.g. \cite{Gaillard:1981rj,Gaillard:1997rt,Gibbons:1995cv,Gibbons:2000xe,Bunster:2011aw,Boulanger:2003vs}. The dynamical field is now a 1-form gauge field $A$ with Abelian gauge transformation
\be 
\d A=\dd \Lambda\,,
\ee
for a  spacetime dependent scalar parameter $\Lambda$.
The Lagrangian is assumed to be an analytic \emph{algebraic} function of the two independent Lorentz invariant scalars in four dimensions, 
\be \label{Lorentz scalars}
\mathrm{k}(F):=\frac{1}{2}\ast(F\wedge\ast F)=-\frac{1}{4}F_{\m\n}F^{\m\n}\,,\qquad \mathrm{t}(F):=\frac{1}{2}\ast(F\wedge F)=\frac{1}{8}\vare^{\m\n\rho\s}F_{\m\n}F_{\rho\s}\,,
\ee
formed by the gauge invariant (constrained) $2$-form field strength $F=\dd A$. The first invariant $\mathrm{k}$ is the U(1) kinetic term, while $\mathrm{t}$ is the (Hodge dual of the) second Chern class giving the topological electromagnetic $\theta$-term. Despite the fact that it is a total derivative (at least when its coupling is a spacetime constant), a generic algebraic function of $\mathrm{t}$ is not. 
For a general such action functional  
\be 
S[A]=\int\mathcal{L}[\mathrm{k}(F),\mathrm{t}(F)]\,\sqrt{|g|}\,\dd^4\s\,,
\ee 
 where $g$ is the determinant of the metric on the 4D spacetime $\S_4$ with Lorentzian signature, the Euler-Lagrange equations read as
\be 
\ast\dd\ast\left(\partial_{\mathrm k}\mathcal{L}\,F-\partial_{\,\mathrm t}\mathcal{L}\,\ast F\right)=0\,,
\ee
where $\partial_{\mathrm k}\equiv \partial/\partial \mathrm k$ and $\partial_{\,\mathrm t}\equiv \partial/\partial \mathrm t$ denote partial differentiation w.r.t. the invariants \eqref{Lorentz scalars}---denoted with Roman typestyle to avoid confusion with indices. In this setting, the quantities $\partial_{\mathrm k/\mathrm t}\mathcal{L}$ should be thought of as scalar functions of $\mathrm k$ and $\mathrm t$. 

Typically, these theories exhibit (generalized) global symmetries with 2-form conserved currents. Indeed, one may directly observe that the following is always such a current: 
\be 
J_{\text{ele}}=\partial_{\mathrm k}\mathcal{L}\,F-\partial_{\,\mathrm t}\mathcal{L}\,\ast F\,,
\ee
 related to the ``electric'' $1$-form (generalized) global symmetry $\d A=a$. Its conservation $\dd\ast J_{\text{ele}}=0$ follows from the field equation of the theory. 
In addition, the Bianchi identity $\dd F=0$ reveals that the ``magnetic'' conserved 2-form current is 
\be 
J_{\text{mag}}=\ast F\,.
\ee
Since both currents are 2-forms,{\footnote{Currents of this sort with equal degree are a feature of theories in dimensions where fields are self-dual, i.e. $2p+2$ dimensions for a $p$-form field. In other dimensions the corresponding currents will be of different degree, cf. \cite{Gaiotto:2014kfa}}} one may couple background $2$-form fields $B$ and $\widehat{B}$ to the theory. This can be done using minimal coupling and one obtains the action
\be \label{action 1 forms + bg}
S[A,B,\widehat{B}]=\int\mathcal{L}[\mathrm k(F,B),\mathrm t(F,B)]\,\sqrt{|g|}\,\dd^4\s-\int_{\S_4}\widehat{B}\wedge F\,,
\ee
where we have defined 
\be \begin{split}
\mathrm k(F,B)&:=\frac{1}{2}\ast[(F-B)\wedge\ast (F-B)]\quad\xrightarrow[]{B\to0}\quad\mathrm k(F)\,,\\ \mathrm t(F,B)&:=\frac{1}{2}\ast[(F-B)\wedge(F-B)]\quad\xrightarrow[]{B\to0}\quad\mathrm t(F)\,.
\end{split}
\ee
As expected, assuming that $B$ and $\widehat{B}$ transform as $\d B=\dd a$ and $\d\widehat{B}=\dd\widehat{a}$, the action \eqref{action 1 forms + bg} fails to be invariant under these background gauge transformations only due to  a $c$-number shift,
\be \label{anomaly 1-forms}
S[A+a,B+\dd a,\widehat{B}+\dd\widehat{a}]-S[A,B,\widehat{B}]=- \int_{\S_4}a\wedge \dd \widehat{B}\,,
\ee
up to boundary terms. This is the 't Hooft anomaly and it can be obtained via inflow from the $6$-form anomaly polynomial 
\be 
\mathcal{I}_6=-\int_{\S_6}\dd B\wedge \dd\widehat{B}\,.
\ee
This was discussed for ordinary electromagnetism in \cite{Cordova:2018cvg}, which in the present setting is obtained simply by assuming that the Lagrangian is equal to $\mathrm{k}$, in which case $\partial_{\mathrm{k}}{\cal L}=1$ and  $\partial_{\mathrm{t}}{\cal L}=0$ and the electric conserved current is given by the field strength $F$ itself. However, as we already observed, this is a feature of any nonlinear electromagnetic theory with duality rotations.

\subsection{Parent action \& duality}
\label{sec32}

In the same philosophy as Section \ref{sec2}, dualization at the off-shell level can be carried out through a parent action functional, which in the present case becomes
\be \label{parent 1-forms}
\mathcal{S}[F,\widehat{A},B,\widehat{B}]=\int\mathcal{L}[\mathrm k(F,B),\mathrm t(F,B)]\,\sqrt{|g|}\,\dd^4\s+\int_{\S_4} F\wedge (\dd\widehat{A}-\widehat{B})\,,
\ee
given in terms of the unconstrained $2$-form $F$, a Lagrange multiplier $1$-form $\widehat{A}$ and the two background fields $B$ and $\widehat{B}$.
One can then integrate out the Lagrange multiplier $\widehat{A}$ and obtain the original second order theory \eqref{action 1 forms + bg} for the constrained $2$-form $F=\dd A$, or integrate out $F$ via the resulting duality relation
\be \label{duality relation 1 forms}
\partial_{\,\mathrm k}\mathcal{L}\, (F-B)-\partial_{\,\mathrm t}\mathcal{L}\,\ast (F-B)=-\ast\,(\dd\widehat{A}-\widehat{B})\,.
\ee
The general procedure would then require us to solve this equation, substitute the solution back into \eqref{parent 1-forms} and obtain a dual second order theory in terms of the $1$-form field $\widehat{A}$ and the background fields. This can be done easily in the linear case of Maxwell's theory, since then $\partial_{\,\mathrm k}\mathcal{L}=1$ and $\partial_{\,\mathrm t}\mathcal{L}=0$. However, when one deals with a general nonlinear theory the aforementioned task can be very demanding, if even possible. 

To complete our set-up, we define the following scalar combinations
\be \label{scalars}\begin{split}
&\ell(F,\widehat{A},B,\widehat{B}):=-\ast\,[(F-B)\wedge \ast(\dd\widehat{A}-\widehat{B})]\,,\\
&\widetilde{\ell}(F,\widehat{A},B,\widehat{B}):=\ast\,[(F-B)\wedge (\dd\widehat{A}-\widehat{B})]\,,\\
&\widehat{\mathrm{k}}(\widehat{A},\widehat{B}):=\frac{1}{2}\ast[(\dd \widehat A-\widehat{B})\wedge \ast(\dd\widehat{A}-\widehat{B})]\,,\\
&\widehat{\mathrm{t}}(\widehat{A},\widehat{B}):=\frac{1}{2}\ast[(\dd \widehat A-\widehat{B})\wedge(\dd\widehat{A}-\widehat{B})]\,.
\end{split}
\ee
The idea now is to multiply the duality relation \eqref{duality relation 1 forms} with appropriate $2$-forms and, subsequently, to act with the Hodge star operator $\ast$ so that \eqref{duality relation 1 forms} gets rewritten as a $4\times 4$ system of equations involving the scalar combinations. Following this recipe, it is straightforward to show that one obtains the system of equations
\be  \label{system of equations}\begin{split}
    \partial_{\,\mathrm{k}}\mathcal{L}\,\mathrm{k}+\partial_{\,\mathrm{t}}\mathcal{L}\,\mathrm{t}&=\frac{1}{2}\widetilde{\ell}\,,\\
    \partial_{\,\mathrm{k}}\mathcal{L}\,\mathrm{t}-\partial_{\,\mathrm{t}}\mathcal{L}\,\mathrm{k}&=\frac{1}{2}\ell\,,\\
    \partial_{\,\mathrm{k}}\,\mathcal{L}\,\ell-\partial_{\,\mathrm{t}}\mathcal{L}\,\widetilde{\ell}&=-2\widehat{\mathrm{t}}\,,\\
    \partial_{\,\mathrm{k}}\,\mathcal{L}\,\widetilde{\ell}+\partial_{\,\mathrm{t}}\mathcal{L}\,\ell&=-2\widehat{\mathrm{k}}\,.
\end{split}
\ee
After some algebra, we observe that the scalar $\ell$ decouples from this system, which was expected since it does not appear in the parent action \eqref{parent 1-forms}. Integrating $\ell$ out and using the first equation to integrate out $\widetilde{\ell}$ from the third and fourth equations, we obtain the equivalent reduced $3\times 3$ system 
\be  \label{final 3x3 system}
\begin{split}
    \partial_{\,\mathrm{k}}\mathcal{L}\,\mathrm{k}+\partial_{\,\mathrm{t}}\mathcal{L}\,\mathrm{t}&=\frac{1}{2}\widetilde{\ell}\,,\\
    \left[(\partial_{\,\mathrm{k}}\mathcal{L})^2-(\partial_{\,\mathrm{t}}\mathcal{L})^2\right]\mathrm{t}-2\,\partial_{\,\mathrm{k}}\mathcal{L}\,\partial_{\,\mathrm{t}}\mathcal{L}\,\mathrm{k}&=-\widehat{\mathrm{t}}\,,\\
    \left[(\partial_{\,\mathrm{k}}\mathcal{L})^2-(\partial_{\,\mathrm{t}}\mathcal{L})^2\right]\mathrm{k}+2\,\partial_{\,\mathrm{k}}\mathcal{L}\,\partial_{\,\mathrm{t}}\mathcal{L}\,\mathrm{t}&=-\widehat{\mathrm{k}}\,.
\end{split}
\ee
Let us now make some remarks. A sufficient condition for a theory described by a Lagrangian $\mathcal{L}[\mathrm{k},\mathrm{t}]$ to admit $SO(2)$ electric-magnetic duality rotations, is that the left-hand side of the second equation above should be equal to $\mathrm{t}$ \cite{Gibbons:1995cv,Gibbons:2000xe,Gaillard:1997rt}. Thus, in these cases the second equation trivializes to $\mathrm{t}=-\widehat{\mathrm{t}}$. Since we are interested only in such kind of duality invariant theories, our general goal will be to obtain real solutions $\mathrm{k}=\mathrm{k}(\widehat{\mathrm{k}},\widehat{\mathrm{t}})$ from the third equation. 
Once a solution has been found, one can compute $\widetilde{\ell}=\widetilde{\ell}(\widehat{\mathrm{k}},\widehat{\mathrm{t}})$ from the first equation and, upon substitution into \eqref{parent 1-forms}, obtain the respective dual action\footnote{Note that, using the definition \eqref{scalars} of $\widetilde{\ell}$, the last term in \eqref{parent 1-forms} can be written in the form $$\int_{\S_4} F\wedge (\dd\widehat{A}-\widehat{B})=-\int\widetilde{\ell}\,\sqrt{|g|}\,\dd^4\s+\int_{\S_4}B\wedge(\dd\widehat{A}-\widehat{B})\,,\quad \sqrt{|g|}\,\dd^4\s=\ast\,\mathbb{I}_4\,.$$}.

A second remark regards the case of Lagrangians with no explicit dependence on $\mathrm{t}$. The most basic example is Maxwell's theory (coupled only to the background $2$-form $B$), which is described by the Lagrangian $\mathcal{L}[\mathrm{k}]=\mathrm{k}$. In these cases, one has $\partial_{\,\mathrm{t}}\mathcal{L}=0$ and the system of equations \eqref{final 3x3 system} further simplifies to
\be  \label{reduced pde system}
    \partial_{\,\mathrm{k}}\mathcal{L}\,\mathrm{k}=\frac{1}{2}\widetilde{\ell}\,,\qquad (\partial_{\,\mathrm{k}}\mathcal{L})^2\,\mathrm{t}=-\widehat{\mathrm{t}}\,,\qquad
    (\partial_{\,\mathrm{k}}\mathcal{L})^2\,\mathrm{k}=-\widehat{\mathrm{k}}\,.
\ee
Again, existence of electric-magnetic duality rotations would require that the l.h.s. of the second equation be equal to $\mathrm{t}$, i.e. that $(\partial_{\,\mathrm{k}}\mathcal{L})^2=1$. This indicates that, from this class of Lagrangians, only Maxwell's theory admits such duality rotations.

\subsection{Examples}\label{sec33}

\subsubsection{Maxwell theory}
Before applying this procedure to more involved examples, let us first convince ourselves that it works by considering the simplest one; Maxwell's theory (coupled to the background fields) with action
\be \label{Maxwell action}
S_{\text{Max}}[A,B,\widehat{B}]=\int\mathrm{k}\,\sqrt{|g|}\,\dd^4\s-\int_{\S_4}\widehat{B}\wedge \dd A\,.
\ee 
For this example, the third equation in \eqref{reduced pde system} has the solution $\mathrm{k}(\widehat{\mathrm{k}})=-\widehat{\mathrm{k}}$. Then, the first equation leads to $\widetilde{\ell}(\widehat{\mathrm{k}})=-2\widehat{\mathrm{k}}$ and one can find the dual action directly from the parent one by mere substitution: 
\be \label{dual Maxwell}\begin{split}
S_{\text{Max}}[\widehat{A},B,\widehat{B}]&=\int\left(\mathrm{k}(\widehat{\mathrm{k}})-\widetilde{\ell}(\widehat{\mathrm{k}})\right)\sqrt{|g|}\,\dd^4\s+\int_{\S_4}B\wedge(\dd\widehat{A}-\widehat{B})\\
&=\int\widehat{\mathrm{k}}\,\sqrt{|g|}\,\dd^4\s+\int_{\S_4}B\wedge(\dd\widehat{A}-\widehat{B}) \,.
\end{split}
\ee
This is the expected dual action in terms of the magnetic potential $\widehat{A}$.
Note in particular that both dual actions share the same 't Hooft anomaly \eqref{anomaly 1-forms}, as expected.
One final comment is that this procedure can take into account the inversion of the dimensionless electromagnetic coupling $e$. Had the first term in \eqref{Maxwell action} been $\frac{1}{e^2}\int\mathrm{k}\,\sqrt{|g|}\,\dd^4\s$, we would have found that the first term in \eqref{dual Maxwell} becomes $e^2\int\widehat{\mathrm{k}}\,\sqrt{|g|}\,\dd^4\s$. This analysis is rather easily extended to the case of the electromagnetic $\theta$-term; a complete discussion is found in \cite{Chatzistavrakidis:2020kpx}, see also \cite{Moreno:2021qae} for a path integral derivation.

\subsubsection{Born-Infeld theory}

Born-Infeld theory \cite{Born:1934gh} is arguably the most famous nonlinear electromagnetic theory. It relates to  Maxwell theory in the same way as the Heisenberg model of Section \ref{sec22} relates to the single compact scalar of Section \ref{sec21}. In other words, it has the same global symmetry structure, even though the electric current takes a different form. Therefore one may consider its action directly coupled to background fields,
\be \label{BI lagrangian}
S_{\text{BI}}[A,B,\widehat{B}]=\int\left(-1+\sqrt{1+2\,\mathrm{k}-\mathrm{t}^2}\right)\sqrt{|g|}\,\dd^4\s-\int_{\S_4}\widehat{B}\wedge \dd A\,,
\ee
which is a function of both $\mathrm{k}$ and $\mathrm{t}$. Moreover, it is invariant under SO(2) electric-magnetic duality rotations \cite{Gaillard:1981rj} and off-shell dualization using the parent action approach can be carried out in any number of spacetime dimensions (see, e.g. \cite{Tseytlin:1996it}). The dual theories are described by actions with the same functional form as \eqref{BI lagrangian}, despite the fact that the dual fields are not $1$-forms in general dimensions other than 4. 
Let us now perform the dualization procedure, as described earlier. Note that the second equation in \eqref{final 3x3 system} becomes $\mathrm{t}(\widehat{\mathrm{t}})=-\widehat{\mathrm{t}}$ (as it should) and the remaining equations are solved by
\be 
\label{system BI}\begin{split}
    \widetilde{\ell}(\widehat{\mathrm{k}},\widehat{\mathrm{t}})&=\frac{-2\,\widehat{\mathrm{k}}+2\,\widehat{\mathrm{t}}^2}{\sqrt{1+2\,\widehat{\mathrm{k}}-\widehat{\mathrm{t}}^2}}\,,\\[4pt]
    \mathrm{k}(\widehat{\mathrm{k}},\widehat{\mathrm{t}})&=\frac{\widehat{\mathrm{k}}\,\widehat{\mathrm{t}}^2-\widehat{\mathrm{k}}+2\,\widehat{\mathrm{t}}^2}{1+2\,\widehat{\mathrm{k}}-\widehat{\mathrm{t}}^2}\,.
\end{split}
\ee
Then, simple substitution back into \eqref{parent 1-forms} yields the dual theory 
\be \label{BI dual}
S_{\text{BI}}[\widehat{A},B,\widehat{B}]=\int\left(-1+\sqrt{1+2\,\widehat{\mathrm{k}}-\widehat{\mathrm{t}}^2}\right)\sqrt{g}\,\dd^4\s+\int_{\S_4}\widehat{B}\wedge (\dd \widehat A-\widehat B)\,,
\ee
in terms of the magnetic potential and exhibiting 
the same 't Hooft anomaly as the electric description of the theory, namely \eqref{anomaly 1-forms}.

\subsubsection{ModMax theory}
Another interesting nonlinear electromagnetism was constructed recently in \cite{Bandos:2020jsw}. It corresponds to the weak-field limit of the most general nonlinear extension to Maxwell's electrodynamics that preserves both SO(2) electric-magnetic duality invariance and conformal invariance. Owing to the fact that this modified Maxwell theory (ModMax) also has two conserved currents, it can be coupled to two background $2$-forms. It is described by a one-parameter action of the form
\be \label{ModMax lagrangian}
S_{\text{MM}}[A,B,\widehat{B}]=\int\left(\cosh(\g)\,\mathrm{k}+\sinh(\g)\,\sqrt{\mathrm{k}^2+\mathrm{t}^2} \right)\sqrt{|g|}\,\dd^4\s-\int_{\S_4}\widehat{B}\wedge \dd A\,.
\ee
Nonlinearity is controlled by the real parameter $\g$, which has to satisfy $\g\geq 0$. As explained in \cite{Bandos:2020jsw}, the reason for this restriction is that otherwise the theory admits superluminal plane wave solutions. Clearly, for $\g=0$ one obtains the action of Maxwell's theory \eqref{Maxwell action}. 

Invariance under SO(2) rotations implies that the second equation in \eqref{final 3x3 system} gives $\mathrm{t}(\widehat{\mathrm{t}})=-\widehat{\mathrm{t}}$. Using this, the third equation in \eqref{final 3x3 system} takes the form
\be \label{intermediate}
\cosh(2\g)\,\mathrm{k}+\sinh(2\g)\sqrt{\mathrm{k}^2+\widehat{\mathrm{t}}^2}=-\widehat{\mathrm{k}}\,,
\ee
which, after some manipulation, can be rewritten as the second order polynomial equation 
\be \label{polynomial equation}
\mathrm{k}^2+2\cosh(2\g)\,\widehat{\mathrm{k}}\,\mathrm{k}+\widehat{\mathrm{k}}^2-\sinh^2(2\g)\,\widehat{\mathrm{t}}^2=0\,.
\ee
The discriminant reads as
\be 
\D=4\sinh^2(2\g)\left(\widehat{\mathrm{k}}^2+\widehat{\mathrm{t}}^2\right)\geq0\,.
\ee
For $\g=0$, the ModMax theory reduces to Maxwell's and this discriminant vanishes. Vanishing of the discriminant implies a unique root and, thus, a unique dual theory. When $\g>0$, however, we observe that there exist two distinct real solutions to \eqref{polynomial equation}, namely
\be \label{solutions modmax}
\mathrm{k}_\pm(\widehat{\mathrm{k}},\widehat{\mathrm{t}})=-\cosh(2\g)\,\widehat{\mathrm{k}}\pm\sinh(2\g)\sqrt{\widehat{\mathrm{k}}^2+\widehat{\mathrm{t}}^2}\,.
\ee
For each $\mathrm{k}_\pm$ one obtains a different $\widetilde{\ell}_\pm$ through the first equation in \eqref{final 3x3 system} and, thus, a different dual action. Substitution of $\mathrm{k}_\pm$ and $\widetilde{\ell}_\pm$ into \eqref{parent 1-forms} gives 
\be \label{dual ModMax}
S^{\pm}_{\text{MM}}[\widehat{A},B,\widehat{B}]=-\int\left(\cosh(\g)\,\mathrm{k}_\pm+\sinh(\g)\,\sqrt{\mathrm{k}^2_\pm+\widehat{\mathrm{t}}^2} \right)\sqrt{|g|}\,\dd^4\s+\int_{\S_4}B\wedge(\dd\widehat{A}-\widehat{B})\,.
\ee
Instead of actually plugging the solutions \eqref{solutions modmax} inside this expression, we can observe that they both satisfy equation \eqref{intermediate} by definition, i.e.
\be 
\cosh(2\g)\,\mathrm{k}_\pm+\sinh(2\g)\sqrt{\mathrm{k}^2_\pm+\widehat{\mathrm{t}}^2}=-\widehat{\mathrm{k}}\,.
\ee
For $\g>0$, we can divide both sides of this relation with $\sinh(2\g)$ and obtain 
\be 
\sqrt{\mathrm{k}^2_\pm+\widehat{\mathrm{t}}^2}=-\frac{1}{\sinh(2\g)}\left(\widehat{\mathrm{k}}+\cosh(2\g)\,\mathrm{k}_\pm\right)\,.
\ee
Finally, we can use this relation to rewrite the dual actions \eqref{dual ModMax} as
\be \begin{split}
S^{\pm}_{\text{MM}}[\widehat{A},B,\widehat{B}]&=\frac{1}{2\cosh(\g)}\int\left(\widehat{\mathrm{k}}-\mathrm{k}_\pm\right)\sqrt{|g|}\,\dd^4\s+\int_{\S_4}B\wedge(\dd\widehat{A}-\widehat{B})\\
&\hspace{-0.28cm}\overset{\eqref{solutions modmax}}{=}\int\left(\cosh(\g)\,\widehat{\mathrm{k}}\mp\sinh(\g)\,\sqrt{\widehat{\mathrm{k}}^2+\widehat{\mathrm{t}}^2} \right)\sqrt{|g|}\,\dd^4\s+\int_{\S_4}B\wedge(\dd\widehat{A}-\widehat{B})\,.
\end{split}
\ee
These actions are distinct for any value of $\g>0$. However, note that $S^{+}_{\text{MM}}(\g)=S^{-}_{\text{MM}}(-\g)$ and, thus, $S^{+}_{\text{MM}}$ has exactly the same functional form as $S_{\text{MM}}$ but with negative angle $\g$. As such, it supports superluminal plane wave solutions and can be discarded. The true dual theory is described by the action $S^{-}_{\text{MM}}$.
Once again, the dual descriptions have the same 't Hooft anomaly \eqref{anomaly 1-forms}.

\subsection{Multiple 1-forms \& generalized target space isometries}
\label{sec34} 

An interesting question arising from the general discussion and examples above is whether the geometric explanation of global symmetries in Section \ref{sec23} in terms of jet space isometries can be extended to capture the case of Abelian $1$-forms in 4D.\footnote{The discussion of non-Abelian fields would be different for a number of reasons. Most importantly, it is well known that Yang-mills theory does not have electric-magnetic duality, at least without supersymmetry. The only known example is the Montonen-Olive duality of $N=4$ super Yang-Mills. However, one can consider a non-Abelian group with an Abelian ideal, i.e. the global symmetry group being $G_{\textrm{non-Abelian}}\times U(1)^n$. Since the duality is only a symmetry for the free theory, this means that one decouples those fields along the Abelian directions which also serve as the dualization directions. One should also note that the action discussed below would have to be suitably modified for non-Abelian fields. Finally, it is worth noting that while flavour symmetries can be non-Abelian, higher-form global symmetries are always Abelian \cite{Gaiotto:2014kfa,Cordova:2018cvg}.} Considering multiple 1-forms $A^{i}$, with $i=1,\dots,D$, one may write any of the theories discussed in the previous section as a generalized theory of maps $A:T\S_4\to {\cal M}$, with a generalized target space of dimension $\text{dim} \,{\cal M}=D$. We emphasize that this generalized sigma model perspective is a natural consequence of the fact that e.g. the free photon is the Nambu-Goldstone boson for spontaneously broken 1-form symmetry \cite{Cordova:2018cvg}. Since effective actions for scalar Nambu-Goldstone bosons are nonlinear sigma models, this generalization to higher spin ones is justified. In the spirit of the coset construction, an extension of the method to higher-form symmetries was proposed in \cite{Landry:2021kko}.

Typically, ${\cal M}$ is a graded manifold equipped with coordinates that carry a degree. In the case at hand, coordinates $a^{i}$ on ${\cal M}$ are assigned degree 1 and their pullback through the map $A$ yield the spacetime 1-forms, $A^{\ast}(a^{i})=A^{i}$. The general action is a theory of such maps, with components the spacetime 1-forms $A^{i}=A^{i}_{\m}\dd \s^{\m}$,
\be \label{Asigmamodel}
S[A]=-\frac 12 \int_{\S_4}\left(G_{ij}(A,\dd A)\,\dd A^{i}\w\ast\,\dd A^{j}+B_{ij}(A,\dd A)\,\dd A^{i}\w\dd A^{j}\right)\,. 
\ee 
One should note that the crucial difference to the scalar case in 2D is that the coupling $B_{ij}$ is now a \emph{symmetric} 2-tensor encoding the generalized theta terms of the multifield theory. Another important difference is that in the present case in order to preserve the gauge invariance, the couplings cannot be functions of $A$. However, we will keep the general dependence on $A$, initially imposing only Lorentz invariance and return to gauge invariance at the end. This philosophy allows for theories that do not have gauge invariance, for instance massive gauge bosons as in the Proca action and higher order terms that combine mass and kinetic ones. As one would expect, this is allowed as long as these sectors correspond to spectator fields and no duality is applied to them. However, $G$ and $B$ can also be functions of gauge-invariant combinations of $\dd A$, as in the case of Born-Infeld and ModMax theories that we have already considered. One may then ask what are the general global symmetries in that case and what is their geometric interpretation in the generalized target space or its extension to jet space thereof.

To address these questions, we consider once again a vector bundle $(E,\pi,\Sigma)$ over the 4D spacetime $\Sigma$ and a coordinate system on $E$  given by $(U,u)$ where this time $u=(\sigma^\mu,a^i)$ and $U$ is an open neighbourhood around a section of the tensor product bundle on $E$ around a point $p\in\Sigma$. Along the same line of reasoning as in Section \ref{sec23}, we can define a 1-jet tensor product bundle such that the induced coordinate system $(U',u')$ on $J^1 \pi$ is given by
\begin{align}
U' &=\left\lbrace j^1_p A^i : A^i(p)\in U\right\rbrace,\\
u' &=\left(\sigma^\mu, a^i, b^i\right),
\end{align}
with $b^i$ coordinates of degree two and, with pullbacks being implicit, 
\begin{align}
a^i&:U\to \mathbb{R},\qquad \qquad a^i = \left. A^i\right\rvert_p,\\
b^i&:U'\to \mathbb{R},\qquad \qquad b^i =\left.\frac{\partial A^i}{\partial \sigma^\mu}\right\rvert_p d\sigma^\mu\,.
\end{align}
Furthermore, consider a Lie algebra valued graded vector field $V = V_a t^a$ on $\mathfrak{X}(E)$, i.e.
\begin{equation}
    V_a = \rho^i_a \frac{\partial}{\partial a^i}\,,
\end{equation}
of degree -1.
A prolongation of this vector to a Lie algebra valued graded vector on $\mathfrak{X}(J^1\pi)$ takes the form
\begin{equation}\label{eq:prol-vect-1form}
V_a = \rho^i_a \frac{\partial}{\partial a^i} + \xi^i_a \frac{\partial}{\partial b^i}\,,
\end{equation}
where $\xi^{i}_{a}$ is of degree 1 and has the form
\begin{align}
\xi^i_a =b^j \frac{\partial \rho^i_a}{\partial a^j}\,.\label{xigen}
\end{align}
This is consistent with the degree of $V_a$ being -1, since $\rho^{i}_{a}$ has degree 0 and $\partial/\partial b^{i}$ has degree -2. 
Also we consider the special case where $\rho^i_a$, $\xi^i_a$ are not explicitly functions of $\sigma^\mu$. 
This graded vector field generates a 1-form global shift symmetry under which $A^i$ and $F^i$ transform as
\begin{align}\label{eq:1formshiftsym}
 \delta A^i = \rho^i_a \epsilon^a \quad \text{and} \quad  
 \delta F^i = \xi^i_a \epsilon^a\,,
\end{align}
where $\epsilon^a$ are the 1-form parameters of the global symmetry and $\xi^i_a =F^j \frac{\partial \rho^i_a}{\partial A^j}$.
Therefore, the transformations of the component fields $A^i_\mu$ and $F^i_{\mu\nu}$ are
\begin{align}\label{eq:1formshiftsym2}
 \delta A^i_\mu = \rho^i_{a} \epsilon^a_{\mu} \quad \text{and} \quad 
 \delta F^i_{\mu\nu} = 2 \xi^i_{a[\mu} \epsilon^a_{\nu]}\,,
\end{align}
which has the usual appearance of a 1-form symmetry for $A^i_\mu$ with $\rho^i_a$ components of a Lie algebra valued vector $\rho^i = \rho^i_a t^a$.

Writing the action \eqref{Asigmamodel} explicitly in terms of the fields $A^i$ and $F^i$,
\be \label{AsigmamodelAF}
S[A,F]=-\frac 12 \int_{\S_4}\left(G_{ij}(A,F)\,F^{i}\w\ast\,F^{j}+B_{ij}(A,F)\,F^{i}\w F^{j}\right)\,, 
\ee 
a direct calculation establishes that it is invariant under the global symmetry \eqref{eq:1formshiftsym} or \eqref{eq:1formshiftsym2} generated by the prolongation vector field \eqref{eq:prol-vect-1form} provided that the following invariance conditions hold
\begin{align}
\widehat{\mathcal{L}}_{V_a} G &= 0\,,\label{eq:Lie_G_vec}\\[4pt]
\widehat{\mathcal{L}}_{V_a} B &= 0\,.\label{eq:Lie_B_vec}
\end{align}
Here $\widehat{\mathcal{L}}_{V_a}$ is the jet space graded Lie derivative, given by a generalization of the formula \eqref{jetlie} to capture the  degree of the vector field $V_a$. It is useful to combine the degree due to grading with the (independent of it) vector field degree into a total degree which is the difference of the two. In that case, denoting the total degree by $|\cdot|$, 
\begin{align}
 |\frac{\partial}{\partial a^k}| = -2\,,\quad  |\frac{\partial}{\partial b^k}| = -3\,,\quad |\rho^i_a| = 0\,,\quad |\xi^i_a| = 1.
\end{align}
In general, for  a Lie algebra valued graded vector field $V_a$ of total degree $p$ and $V_1,V_2 \in \mathfrak{X}(J^1 \pi)$ two graded vectors of total degree $k$, we have
\begin{align}
(\widehat{\mathcal{L}}_{V_a} G)(V_1,V_2) = V_a(G(V_1,V_2))
- G(\widehat{\mathcal{L}}_{V_a} V_1,V_2)- \alpha^{k+2}\, G(V_1,\widehat{\mathcal{L}}_{V_a} V_2)\,,\label{jetNijlie}
\end{align}
where $\alpha = +1$ for even $p$ and $\alpha = -1$ for odd $p$.
For the case discussed here, $V_a,V_1,V_2$ are all of total degree $-2$, which results in 
\begin{align}
(\widehat{\mathcal{L}}_{V_a} G)(V_1,V_2) = V_a(G(V_1,V_2))
- G(\widehat{\mathcal{L}}_{V_a} V_1,V_2)-\, G(V_1,\widehat{\mathcal{L}}_{V_a} V_2)\,,\label{jetNijlie2}
\end{align}
The same formula holds for $B$, since it is also a symmetric 2-tensor. 
The couplings $G$ and $B$ of the 4D theory are thus interpreted as generalized target space metrics and the generalized global symmetries of any 4D nonlinear electromagnetism of the type considered here acquire a geometric interpretation as Killing isometries for generalized target space background fields.

Coming back to gauge invariance, the dependence of $G$ and $B$ should be solely on $F$ and not explicit in $A$. This is the case for all theories considered in this section. When the functions $\rho^{i}_a$ that control the global symmetry are also $A$-independent there is shift symmetry $\d A^i=\epsilon^i$ for the 1-form rigid parameter $\epsilon^i$. One may then observe that the invariance conditions are satisfied because due to \eqref{xigen} we obtain that $\xi^{i}_a=0$, which does not exclude the dependence of $G$ and $B$ on $F$. This is precisely what happens in the examples of Born-Infeld and ModMax, where the coupling $G$ depends on $F$, the generalized global symmetry is a constant shift of $A$ and therefore the invariance conditions are satisfied identically.

Finally, we briefly comment on what happens to the couplings $G$ and $B$ under electric/magnetic duality. In the sigma model philosophy, realized here through the action \eqref{Asigmamodel}, we introduce the parent action
\be \label{eq:mult_vec_parent}
{\cal S}[F,\widehat{A}\,]=-\frac 12 \int_{\S_{4}} \left(G_{ij} \,F^i \w\ast F^j + B_{ij}\, F^i \wedge F^j\right)+\int_{\S_4} F^m \w  \dd\widehat{A}_m\,,
\ee 
where $\widehat{A}_m$ is a Lagrange multiplier for the constraint $\dd F^m =0$ and the index $m$ refers to the directions that undergo duality, namely $A^{i}=(A^{\a},A^{m})$ with $A^{\a}$ the spectator fields. Following the same procedure described several times in this paper, we can write down the dual action for the field $\widehat{A}^{i}=(A^{\a},\widehat{A}_m)$,
\begin{align}
\widehat{S}[\widehat{A}\,] =-\frac{1}{2} \int_{\Sigma_4}\, \bigg(\widehat{G}_{ij}\,\dd\widehat{A}^i \wedge  \ast\,\dd\widehat{A}^j + \widehat{B}_{ij}\, \dd\widehat{A}^i \wedge \dd\widehat{A}^j\bigg),
\end{align} 
where $\widehat{G}_{ij},\widehat{B}_{ij}$ are the electromagnetic dual couplings obtained from the original ones through the action of the duality group $Sp(2D,\mathbb{R})$. In order to see this, it is convenient to introduce the complex coupling $\tau_{ij} =B_{ij}+ i G_{ij}$. Upon dualization we will have the duality rules \cite{Chatzistavrakidis:2020kpx}
\be
\widehat{\tau}_{ij} = \tau_{ij} - \tau_{im}\tau^{mn}\tau_{nj},\quad \widehat{\tau}_{im} = -\tau_{in} \tau^{nm},\quad \widehat{\tau}_{mn} = -\tau^{mn},
\ee
which are the fractional linear transformation of $\tau$ under $Sp(2D,\mathbb{R})$ group action. This can be shown to be the case for all sigma models of the fields with odd form degree in self-dual dimensions \cite{Gaillard:1981rj,Bunster:2011aw}.

\section{Discussion \& conclusions}
\label{sec4}

In this work we discussed a variety of relativistic, linear and nonlinear field theories for single or multiple scalar fields and 1-forms that feature duality rotations and ordinary or generalized global symmetries in their self-dual dimensions. We highlighted that several well known properties of the single compact scalar and its multifield nonlinear sigma model extension in 2D hold for such more general theories as well, specifically generalizing  in three directions, namely for
\begin{itemize} 
\item higher derivative scalar field theories in 2D,
\item Abelian 1-form gauge theories in 4D, such as Maxwell, Born-Infeld and ModMax, 
\item 4D gauge theories for multiple Abelian 1-forms.
\end{itemize}
For this general class of theories, we identified the conserved currents of their  global symmetries and by coupling background fields to them discussed how their 't Hooft anomaly and its corresponding anomaly polynomial are derived. 
The features that all these types of theories have in common can be summarized in three main lessons, which constitute the results of our analysis:
\begin{enumerate}
\item Global symmetries (ordinary or generalized) correspond to Killing vector isometries in a suitable (graded) target space of fields. 
\item Theories with global symmetries can be formulated as generalized sigma models.
\item Couplings (target space background fields) transform under duality rotations with one of two sets of Buscher-like rules.
\end{enumerate}
An underlying reason for this universal perspective is that generalized global symmetries can be spontaneously broken and give rise to Nambu-Goldstone bosons which can be higher spin particles, extending the usual scalar Nambu-Goldstone bosons of flavour symmetries \cite{Gaiotto:2014kfa,Cordova:2018cvg}. For instance, the photon is interpreted as the Nambu-Goldstone ``vector'' for a broken 1-form global symmetry---see also \cite{Asakawa:2012px,Gliozzi:2011hj,Gomis:2012ki} for a somewhat different D-brane perspective where this happens due to spontaneous breaking of Lorentz symmetry by the brane. The intimate relation between nonlinear sigma models and Nambu-Goldstone bosons is then reflected in items 1 and 2 above. From a more mathematical standpoint, the origins of this unity can be traced back to the universal formulation of this large class of theories and their dualities in terms of graded geometry \cite{Chatzistavrakidis:2019len}. This also underscores the fact that duality, even T, is not a feature of a specific model but it  exists whenever there are two global symmetries of similar form.

Although we have not discussed in detail $p$-form theories in their self-dual dimensions, the above results readily generalize to them and their $p$-form global symmetries with conserved currents being differential forms of degree $p+1$. Nevertheless, it is worth stressing that as far as duality rules are concerned, even degree forms follow the scalar Buscher rules whereas odd degree ones follow the 1-form Buscher-like rules \cite{Chatzistavrakidis:2020kpx}. In fact, at this level this has been extended even to mixed symmetry tensors of type $(p,1)$ for linear theories with multiple such fields for even or odd $p$. To give an example, linearized gravity described by a symmetric 2-tensor $h_{\m\n}$, the graviton, exhibits an electromagnetic type duality that exchanges the Riemann tensor with its dual and the linearized Einstein equations and algebraic and differential Bianchi identities with the ones for the dual graviton \cite{Hull:2001iu,West:2001as,Bekaert:2002dt,deMedeiros:2002qpr}. Multiple (non-interacting) gravitons can then lead to a set of duality rules with the same form as the electromagnetic ones. Note however that this does not extend to gravity theories of the Born-Infeld type \cite{BeltranJimenez:2017doy} since these lack Lorentz invariance. Nevertheless, it would be interesting to explore the apparent generalized global symmetry of the linearized Einstein-Hilbert action under tensor shifts, namely $\d h_{\m\n}=s_{\m\n}$ for a constant symmetric tensor $s_{\m\n}$, identify the corresponding 't Hooft anomaly and examine whether the graviton could be interpreted as a Nambu-Goldstone boson for this global symmetry.  

Finally, although we have restricted our analysis to self dual dimensions, one could work in general dimensions too. The dual fields in such a case are of different type, for example 1-forms in three dimensions are dual to scalars and consequently such theories have generalized global symmetries of different order. In a different spirit, the generalized sigma model perspective could also be useful to analyze scalar field theories in higher dimensions than two, as for example the higher derivative brane model in four dimensions constructed in 
\cite{Romoli:2021hre}.

\paragraph{Acknowledgements.}  We would like to thank Riccardo Argurio, Larisa Jonke, Carlos N\'u\~nez, Jan Rosseel and Peter Schupp for useful discussions. This work is supported by the Croatian Science Foundation Project ``New Geometries for Gravity and Spacetime'' (IP-2018-01-7615). A. Ch. is grateful to the Erwin Schr\"odinger International Institute for Mathematics and Physics for hospitality and financial support in the framework of the Research in Teams Project: Higher Global Symmetries and Geometry in (non-)relativistic QFTs.

\end{document}